\documentclass [12PT]{article}

\usepackage{float}
\usepackage[comma,sort&compress]{natbib}
\usepackage{multirow}
\usepackage{multicol}
\usepackage{graphicx}
\usepackage{mathtools,amssymb,amsmath}
\usepackage[title]{appendix}
\usepackage{color,bm,xcolor}
\usepackage{url}
\usepackage{setspace}
\usepackage{smartref}
\usepackage{rotating}
\usepackage{algorithm,algorithmic}
\usepackage{mathtools}
\usepackage{bbm}
\usepackage{subfigure}
\DeclarePairedDelimiter{\ceil}{\lceil}{\rceil}
\DeclarePairedDelimiter\floor{\lfloor}{\rfloor}

\def \bSig \mathbf{\Sigma}

\newcommand{\bb}[1]{\boldsymbol{#1}}
\providecommand{\keywords}[1]{\textbf{\textit{Keywords:}} #1}
\newcommand{\bY}{{\bb Y}}

\newcommand{\bp}{{\bm p}}

\definecolor{ying}{rgb}{0.123,0.545,0.123}

\usepackage{fancyhdr}
 {
\fancyhf{}
\fancyhead[LE]{Lyu \textit{et. al.}}
\fancyhead[RO]{Bayesian Sample Size Determination}
\fancyfoot[C]{\thepage}

}

\begin{document}

\title{BaySize: Bayesian Sample Size Planning for Phase I Dose-Finding Trials} 
\author{Xiaolei Lin$^1$, Jiaying Lyu$^2$, Shijie Yuan$^2$, Sue-Jane Wang$^3$, Yuan Ji$^{4,*}$ \\ 
	$^1$ \small School of Data Science, Fudan University, Shanghai, China\\
	$^2$ \small 1. Laiya Consulting, Inc., Shanghai ,China\\	
	$^3$ \small US Food and Drug Administration, Silver Spring USA. \\    
	$^4$ \small Department of Public Health Sciences, The University of Chicago, Chicago, USA\\	    
	$^*$ \small corresponding: koaeraser@gmail.com;
}
\date{}           
\maketitle

\begin{abstract}
We propose BaySize, a sample size calculator for phase I clinical
trials using Bayesian models. BaySize applies the concept of effect
size in dose finding, assuming the MTD is defined based on an
equivalence interval.   Leveraging a decision framework that
involves composite hypotheses, BaySize   utilizes   two prior
distributions,   the fitting prior (for model fitting) and sampling
prior (for data generation),   to conduct sample size calculation under
desirable statistical power.   Look-up tables are generated to facilitate practical applications. To our knowledge, BaySize is the first sample size tool that can be applied to a broad range of phase I trial designs.

\end{abstract}

\section{Introduction}
The primary objective of a phase I dose-finding trial is to identify the maximum tolerated dose (MTD) of an experimental drug. MTD is typically defined as the highest dose where the probability of the dose limiting toxicity (DLT) is close to or no higher than the target toxicity rate $p_T$, say $p_T=0.30$. During the course of a trial, patients are accrued sequentially in cohorts, and dose assignment to a cohort of patients is dependent upon observed toxicity outcomes of the previously treated cohorts. Over the past three decades, a large number of dose-finding approaches have been proposed, including, for example, the 3+3 design \citep{storer1989design}, the continual reassessment method (CRM) \citep{o1990continual}, the escalation with overdose control (EWOC) design \citep{babb1998cancer}, the cumulative cohort (CCD) design \citep{ivanova2007cumulative}, the Bayesian optimal interval (BOIN) design \citep{liu2015bayesian}, the modified toxicity probability interval (mTPI, mTPI-2) methods \citep{ji2010modified,ji2013modified,guo2017bayesian}, and the i3+3 design \citep{i3plus3}.

\vskip 0.1in
\noindent
The number of patients required to achieve the scientific goal of a clinical trial is a major task to be considered in a statistical design since investigators only possess limited amount of resources and time. Larger sample sizes increase the power of achieving trial objectives but also increase trial cost and duration. Despite flourishing literature in novel dose-finding designs, only a limited number of methods has been proposed for sample size determination for phase I trial design. \\

\vskip 0.1in
\noindent
For clinical trials in general, not necessarily Phase I trials, Bayesian sample size determination approaches are based on various performance metrics, including, for example, the average posterior variance criterion (APVC) \citep{wang2002simulation}, the average length criterion (ALC), the worst outcome criterion \citep{joseph1995sample,joseph1997bayesian,m2006bayesian,m2008bayesian}, and Bayes factors \citep{weiss1997bayesian}. \citet{zohar2001continual} calculated the required number of patients by considering different stopping rules in the CRM design. \citet{lin2001statistical} and \citet{ivanova2006escalation} described sample size recommendations on the basis of the expected number of patients allocated to each dose among prespecified dose levels. \citet{tighiouart2012number} estimated the sample size using the posterior standard deviation and the highest posterior density interval of the MTD in the EWOC design. Recently, \citet{kuen2013sample} presented a closed-form formula for sample size calculation, in which a nonparametric optimal design \citep{o2002non} was exploited and the accuracy index of the CRM design was derived empirically.  These approaches, however, are tailored for specific designs, such as CRM or EWOC. \\

\vskip 0.1in
\noindent
In this paper, we propose a general   decision   framework for
sample size determination based on Bayesian   hierarchical
modeling. We set up two composite hypotheses indicating the presence
and location of the true MTD, and apply Bayesian decision rules to
select one of the hypotheses. Sample size is therefore decided
according to desirable statistical power.   
Specifically, BaySize defines a   composite   null hypothesis
$H_0$ ``none of the doses are the MTD'' and   a composite  
alternative hypothesis $H_1$ ``one of the doses is the MTD'',   and
powers dose-finding trials based on the power of a Bayesian decision
using the Bayes factor (BF).    
In addition, the proposed BaySize framework incorporates two types of
priors - the fitting prior for BF estimation and the sampling prior
for trial data generation,   in order   to achieve robustness and avoid inflating Type I error rate \citep{li2000bayesian}.  \\

\vskip 0.1in
\noindent
The remainder of the article is organized as follows. Section \ref{Sec:bssd} illustrates the BaySize framework in details and provides a searching algorithm for the desired sample size. A trial example is given  in Section \ref{Sec:application}. In Section \ref{Sec:sim}, the performance of BaySize is evaluated via extensive simulation studies with sensitivity analyses.  The paper closes with a brief discussion in Section \ref{Sec:dis}. \\

\section{The BaySize Framework}\label{Sec:bssd}

\subsection{Bayesian Hypothesis Testing for Phase I Dose-Finding Trials}\label{Sec:bht}

Recent interval-based dose-finding designs, such as BOIN, i3+3, mTPI-2 and SPCRM \citep{liu2015bayesian, i3plus3, guo2017bayesian, clertant2017semiparametric} define MTD as a dose with a toxicity probability falling into the EI, where EI$=[p_T - \epsilon_1, p+T + \epsilon_2]$ uses two small fractions. Suppose a total of $D$ dose levels are prespecified for a Phase I dose-finding trial. Consider two hypotheses: 

\begin{equation*}
H_0: \mbox{none of the prespecified doses are MTD} 
\quad \mbox{v.s.} \quad 
H_1: \mbox{ one of the prespecified doses is MTD}.
\end{equation*} 

\noindent
Denote $p_d$ the DLT probability of dose $d$, $d=1,2,\ldots,D$.   We
consider two composite  hypotheses regarding the presence and
location of the true MTD.  

\begin{equation}
H_0: \forall \; d, \; p_d \notin [p_T-\epsilon_1,p_T+\epsilon_2]  
\quad \quad \mbox{v.s.} \quad \quad
H_1: \exists \; d, \; p_d \in (p_T-\epsilon_1,p_T+\epsilon_2). \label{hypoth}
\end{equation}

\noindent
We define $(\epsilon_1 + \epsilon_2)$ as the ``effect size'' of the design, which describes the desired precision in MTD identification. The EI can be easily elicited from physicians by asking them the lowest and highest percentages of patients who have DLTs among, say 100 patients treated at the MTD. The effect size quantifies the desired accuracy in MTD identification. The larger or smaller the value of $(\epsilon_1 + \epsilon_2)$, the less or more accurate the MTD is to be identified, respectively. Since effect size is defined through the specification of EI in the context of interval-based designs, we hereinafter use the mTPI-2 design \citep{guo2017bayesian} as an example to further elaborate the BaySize framework. A brief discussion of the trivial extension of BaySize to accommodate other dose-finding designs is provided at the end of the paper. \\

\vskip 0.1in
\noindent
Let $X_d$ be the number of patients with DLTs and $N_d$ the total number of patients treated at dose $d$, respectively, for $d=1,2,\ldots,D$. Denote $\bp=(p_1,p_2,\ldots,p_D)$ the vector of  the unknown toxicity probability at the $D$ doses and $\bY \equiv \{(X_d,N_d),d=1,2,\ldots,D\}$ the observed trial data. Under Bayesian hypothesis testing, a prior distribution of $\bp$ is constructed under each of the two hypotheses, $H_0$ and $H_1$, defined as  $\pi^{(f)}_0(\bp|H_0)$ and $\pi^{(f)}_1(\bp|H_1)$, respectively. Here the superscript ``$f$'' implies that these are the priors used to fit the data. Therefore, we call $\pi^{(f)}_0(\bp|H_0)$ and $\pi^{(f)}_1(\bp|H_1)$ the ``fitting priors''. Later we will introduce another pair of priors called ``sampling priors'', which are assumed to be the generative model for the trial data. The use of fitting and sampling priors is originally proposed in \citep{li2000bayesian} for non-inferiority trials. Given a dose-finding design $\mathcal{E}$ (say, the mTPI-2 design) and the vector of toxicity probability $\bp$, trial data $\bY$ are assumed to follow a sampling distribution $f(\bY|\bp, \mathcal{E}) $. For the ease of exposition, hereinafter we drop the dependence on  $\mathcal{E}$  in mathematical derivation unless otherwise stated. For example, we will write $f(\bY|\bp)$ instead of $f(\bY|\bp,  \mathcal{E} )$. But keep in mind that the subsequent model development always assumes an underlying dose-finding design  $\mathcal{E}$. \\

\vskip 0.2in
\noindent
The Bayesian hypothesis testing is based on the Bayes Factor $B(\bY)$, which is defined as 

\begin{equation}
B(\bY)=\frac{f(\bY|H_0)}{f(\bY|H_1)}=\frac{\int f(\bY|\bp) \cdot
  \pi_0^{(f)}(\bp|H_0)  d\bp}{\int f(\bY|\bp) \cdot
  \pi_1^{(f)}(\bp|H_1)  d\bp} \label{BF}
\end{equation}

\noindent
Given a critical cutoff value $BF_0$, the Bayesian test chooses $H_1$ if $B(\bY)<BF_0$,  and $H_0$ otherwise. \\

\subsection{Sample Size Determination}\label{Sec:ssd}
For a Type I error rate $\alpha$ and Type II error rate $\beta$,  BaySize determines the trial sample size by seeking an integer $n$ that satisfies 

\begin{eqnarray*}
	\xi_0(n) \equiv Pr \left\{ B(\bY^{(n)}) < BF_0|H_0 \right\} & \leq & \alpha  \quad \mbox{(Type
  I error)}, \quad \mbox{and} \\
	\xi_1(n) \equiv Pr\left\{ B(\bY^{(n)})<BF_0|H_1 \right \} & \geq & 1-\beta  \quad \mbox{(Power)},
\end{eqnarray*}

\noindent
where $\bY^{(n)}$  denotes the trial data under a sample size $n$. The two probabilities $\xi_0(n)$ and $\xi_1(n)$ are evaluated under the marginal distribution $f(\bY^{(n)} | H_i)$ defined as

\begin{equation}
f(\bY^{(n)} | H_i) = \int f(\bY^{(n)}|\bp) \cdot \pi^{(s)}_i(\bp|H_i) \ d\bp, \quad \quad i=0, 1,
\end{equation}

\noindent
where $\pi^{(s)}_i(\bp|H_i)$ is the sampling prior that generates the observed trial data $\bY^{(n)}$. Also $f(\bY^{(n)} | H_i)$ is known to Bayesians as the prior predictive distribution. Therefore, by definition we have 

\begin{eqnarray}
\xi_{0}(n)  &=& \iint
\bm{1}\left\{B(\bY^{(n)})<BF_0\right\}f(\bY^{(n)}|\bp)\pi^{(s)}_0(\bp|H_0) \;d\bp\;d\bY^{(n)} \label{error}, \quad \mbox{and}\\
\xi_{1}(n)  &=& \iint
\bm{1}\left\{B(\bY^{(n)})<BF_0\right\}f(\bY^{(n)}|\bp)\pi^{(s)}_1(\bp|H_1) \;d\bp\;d\bY^{(n)} \label{power}
\end{eqnarray} 

\noindent
where $\bm{1}\{\cdot\}$ is the indicator function that takes values 0 or 1. For sample size determination, the sampling priors $\pi^{(s)}$ and fitting priors $\pi^{(f)}$ should be different in order to avoid inflating the Type I error rate \citep{li2000bayesian}. More details about both priors are provided in Sections \ref{Sec:fp} and \ref{Sec:sp}. Since sample size determination must be done before the trial starts, the trial data $\bY^{(n)}$ has not yet been observed and is assumed to be random. Therefore, the Bayes Factor $B(\bY^{(n)})$ and the decision rule $\left \{B(\bY^{(n)})<BF_0\right\}$, which are functions of the trial data $\bY^{(n)}$, are also random. 

\vskip 0.1in
\noindent
For a pre-determined Type I error rate $\alpha$ and power $(1-\beta)$, BaySize determines the sample size by calculating 

\begin{equation}
\label{box}
\boxed
{n^\star=min\{integer \quad n: \xi_0(n) \le \alpha \quad and \quad \xi_1(n) \ge 1-\beta \}} 
\end{equation}

\noindent
The main question is how to calculate $n^\star$, which will be explained in the remainder of Section \ref{Sec:bssd}.

\subsection{The Fitting Prior $\pi_0^{(f)}$ and $\pi_1^{(f)}$} \label{Sec:fp}
Both the fitting and sampling priors must be specified in order to calculate $B(\bY^{(n)})$ and evaluate $\xi_0(n)$ and $\xi_1(n)$. We adopt the construction of the fitting priors $\pi^{(f)}(\bp \mid H_0)$ and $\pi_1^{(f)}(\bp \mid H_1)$ in the semi-parametric CRM model (SPCRM) by \cite{clertant2017semiparametric}. \\

\vskip 0.1in
\noindent
Under $H_1$, the model space in SPCRM is partitioned by $D$ submodels that provide the exact true MTD locations, given by
\begin{eqnarray*}
	&M_{11}:& p_1 \in (p_T-\epsilon_1,p_T+\epsilon_2), \mbox{~~and~~} \forall \ d \neq 1, \; p_d \notin (p_T-\epsilon_1,p_T+\epsilon_2);\\
	&M_{12}:& p_2 \in (p_T-\epsilon_1,p_T+\epsilon_2), \mbox{~~and~~} \forall \ d \neq 2, \; p_d \notin (p_T-\epsilon_1,p_T+\epsilon_2);\\
	&\cdots& \\
	&M_{1D}:& p_D \in (p_T-\epsilon_1,p_T+\epsilon_2), \mbox{~~and~~} \forall \ d \neq D, \; p_d \notin (p_T-\epsilon_1,p_T+\epsilon_2). 
\end{eqnarray*} 

\noindent
It is trivial that $H1 = \bigcup_{d=1}^D M_{1d}$, since each submodel $M_{1d}$ specifies one and only one dose level $d$ as the true MTD. For example, submodel $M_{11}$  states that only the first dose is in the equivalence interval and hence is the only true MTD.  Given each submodel, the fitting prior can be easily specified according to \citet{clertant2017semiparametric}. Specifically, conditional on model $M_{1d}$, the fitting prior $\pi^{(f)}_{1d}(\bp|M_{1d})$ takes the form of a product of the independent truncated beta distribution, given by 

\begin{eqnarray}
  \pi^{(f)}_{1d}(\bp |M_{1d}) &\propto&
\left[\prod_{k=1}^{d-1}\mathcal{B}_{LI}\left\{cq^{1d}_k+1,c(1-q^{1d}_k)+1\right\} 
\right] \cdot \mathbbm{I}(d>1)
\times \nonumber \\
&& \mathcal{B}_{EI}\left\{cq^{1d}_d+1,c(1-q^{1d}_d)+1\right\}
\times
\left[\prod_{k=d+1}^{D}\mathcal{B}_{HI}\left\{cq^{1d}_k+1,c(1-q^{1d}_k)+1\right\} \right] \cdot \mathbbm{I}(d<D), 
\label{eq:beta1} 
\end{eqnarray}

\noindent
where $\mathcal{B}_{LI}$, $\mathcal{B}_{EI}$ and $\mathcal{B}_{HI}$ denote the truncated beta distributions on the lower interval (LI) $(0,p_T-\epsilon_1)$, the equivalence interval (EI) $[p_T-\epsilon_1,p_T+\epsilon_2]$ and the higher interval (HI) $(p_T+\epsilon_2,1)$, respectively, $c \ge 0$ is a dispersion parameter, and $q_k^{1d}$ is the mode of the truncated prior for the $k$th dose under equation \eqref{eq:beta1}. Let $\bm{q^{1d}}=(q^{1d}_1,\cdots,q^{1d}_D) \in [0,1]^D$ be the vector of the modes. A simple specification of $\bm{q^{1d}}$ is provided in Appendix \ref{modes} extending the idea in \citep{clertant2017semiparametric}. 

\vskip 0.1in
\noindent
Similarly, under $H_0$, the model space is partitioned into submodels as follows: 

\begin{eqnarray*}
	&M_{00}:& \forall \;d, \; p_d \in (p_T+\epsilon_2,1); \\
	&M_{01}:& p_1 \in (0,p_T-\epsilon_1) \ \mbox{~~and~~} \forall\ d>1,\; p_d \in (p_T+\epsilon_2,1); \\
	&M_{02}:& \forall\ d \leq 2, \; p_d \in (0,p_T-\epsilon_1) \ \mbox{~~and~~} \forall\ d>2,\; p_d \in (p_T+\epsilon_2,1); \\
	&\cdots& \\
	&M_{0,D-1}:& \forall\ d \leq D-1, \ p_d \in (0,p_T-\epsilon_1), \ \mbox{~~and~~} p_D \in (p_T+\epsilon_2,1);\\
	&M_{0D}: & \forall \;d, \; p_d \in (0,p_T-\epsilon_1). 
\end{eqnarray*}

\noindent
Each submodel corresponds to a subspace where none of the $D$ doses are the true MTD. Specifically, the first model $M_{00}$ indicates that all the doses are higher than the true MTD, the last model ($M_{0D}$) indicates that all the doses are lower than the true MTD, and the remaining models in between imply that the true MTD is sandwiched by two adjacent doses.  Conditional on model $M_{0d}$, the fitting prior is given by

\begin{eqnarray}   
\pi^{(f)}_{0d}(\bp|M_{0d}) \propto
 \left[ \prod_{k=1}^{d}\mathcal{B}_{LI}\left\{cq^{0d}_k+1,c(1-q^{0d}_k)+1\right\} \right] \cdot \mathbb{I}(d>0)
\times 
\left[\prod_{k=d+1}^{D}\mathcal{B}_{HI}\left\{cq^{0d}_k+1,c(1-q^{0d}_k)+1\right\}\right] \mathbb{I}(d<D). \label{eq:beta2}
\end{eqnarray} 

\noindent
where $\mathcal{B}_{LI}$, $\mathcal{B}_{HI}$, $q_{0d} = (q_0^{0d}, ..., q_D^{0d})$ and $c$ denote the truncated beta distributions on the lower and higher intervals, their corresponding mode vector, and dispersion parameter, respectively. \\

\vskip 0.1in
\noindent
To calculate the Bayes factor in \eqref{BF}, the following hierarchical model is specified. Denote $m$ as the model indicator, for $i=0,1$, 

\begin{eqnarray*}
	\bY^{(n)} \mid \bp,  \mathcal{E} &\sim&   f(\bY^{(n)} | \bp,  \mathcal{E}), \\ 
	 \pi_{id}^{(f)}(\bp \mid  m=M_{id})  &\sim& \mbox{ \eqref{eq:beta1} or \eqref{eq:beta2}}, \qquad i=1,2, \quad d=1, ..., D \\
	 M_{id} \mid H_i  &\sim& \pi(M_{id}\mid H_i)\cdot \mathbbm{I}\{M_{id} \in \mathcal{M}_i\}, \qquad i=1,2, \quad d=1, ..., D \\
	 Pr(H_i) &=& \frac{1}{2}, \quad i=1,2
\end{eqnarray*}

\noindent
where the sets $\mathcal{M}_1=\{M_{11},M_{12},\ldots,M_{1D}\}$ and $\mathcal{M}_0=\{M_{00},M_{01},\ldots,M_{0D}\}$ contain $D$ and $(D+1)$ models, respectively. Without loss of generality, $\pi(M_{id}|H_i)$ is assumed to be uniform among the models in $\mathcal{M}_0$ and $\mathcal{M}_1$, i.e., each model in $\mathcal{M}_0$ (or $\mathcal{M}_1$) is weighted equally under $H_0$ (or $H_1$). However, other prior can be considered based on available information for the doses. For example, one may assume $Pr(m=M_{1d} \mid H_1)$ is small for $d=1$. That is, it is unlikely that the lowest dose is the MTD due to the fact that the lowest dose in a practical dose-finding trial is usually below the therapeutic window for safety reasons. Note that under the above hierarchical model specification, the fitting priors under $H_0$ and $H_1$ can be derived by integrating out $m$, i.e., $\pi^{(f)}_i(\bp|H_i)=\sum_{M_{ij} \in \mathcal{M}_i}
\pi^{(f)}_{id}(\bp|m=M_{id},H_i) Pr(m=M_{id}|H_i)$, for $i=0,1$.

\subsection{The Sampling Prior $\pi_0^{(s)}$ and $\pi_1^{(s)}$} \label{Sec:sp}

Box \eqref{box} requires the evaluation of $\xi_0(n)$ and $\xi_1(n)$, which will be conducted by numerical simulation. Specifically, $\bY^{(n)}$ will be generated from the prior predictive distribution involving the construction of priors. To avoid inflating the Type I error rate and overfitting, the BaySize generates trial data using the sampling priors $\pi_0^{(s)}(\bp \mid H_0)$ under $H_0$ and $\pi_1^{(s)}(\bp \mid H_1)$ under $H_1$, which differ from the aforementioned fitting priors. In this section, we provide some examples for sampling prior specification. \\

\vskip 0.1in
\noindent
In practice, a statistical design $\mathcal{E}$ is typically evaluated by simulating trial data under various scenarios. True probabilities of the candidate doses are prespecified for each scenario. Let $\bp_1^*$ denote the prespecified toxicity probabilities among which only one of the toxicity probabilities falls in the EI (i.e., one of the dose levels can be considered as the MTD). For example, if $p_T=0.3$, EI$=[0.2,0.4]$, and $D=5$, then $\bp_1^*$ may be set as $(0.05, 0.09, 0.12, 0.19, 0.25)$. One simple method is to set $\pi_1^{(s)}(\bp \mid H_1)$ as the point mass at $\bp_1^*$, given by
 
\begin{equation}
\pi^{(s)}_1(\bp|H_1)=\bm{1}\{\bp=\bp_1^*\}
\label{eq:sp.h1}
\end{equation} 

\begin{figure}
	\centering \includegraphics[width=0.7\textwidth]{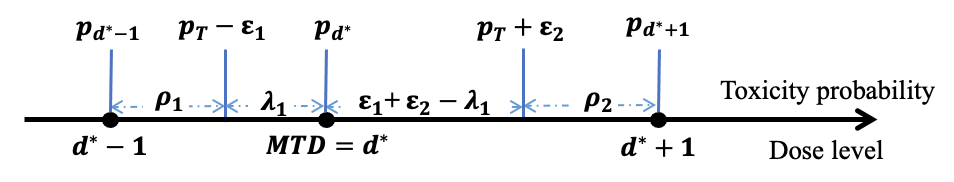}
	\caption{Factors that could potentially impact the configuration of $\bp_1^*$ and the associated sampling prior in \eqref{eq:sp.h1}: the target toxicity rate ($p_T$), the effect size ($\epsilon_1$ and $\epsilon_2$), the MTD location in the dose range (i.e., the true MTD level, $d^*$), the MTD location in the EI (i.e., the distance from the lower boundary of EI $p_T-\epsilon_1$ to the MTD $d^*$, $\lambda_1$), and the distances from the lower / upper boundary of the EI to the next doses below / above the MTD ($\rho_1$ and $\rho_2$).}
	\label{fig:sp1}
\end{figure}

\vskip 0.1in
\noindent
Different $\bp_1^*$ values yield different sample size calculation. For instance, when the target toxicity probability $p_T = 0.3$ and EI $= [0.2, 0.4]$ (i.e., $\epsilon_1=\epsilon_2=0.1$), larger sample size is required for $\bp_1^* = (0.05, 0.09, 0.12, 0.19, 0.25)$ compared with $\bp_1^* = (0.01, 0.3, 0.5, 0.6, 0.7)$, because the MTD is easier to identify for the latter case $\bp_1^* = (0.01, 0.3, 0.5, 0.6, 0.7)$. This is simply because escalation to dose level 2 in $\bp_1^* = (0.01, 0.3, 0.5, 0.6, 0.7)$ would achieve the MTD while in the former case the trial will need to escalate all the way to dose level 3 or 4. In addition to $p_T$ and effect size $(\epsilon_1 + \epsilon_2)$, several other factors need to be considered in specifying $\pi_1^{(s)}(\bp \mid H_1)$ as reflected in Figure \ref{fig:sp1}: 1) relative MTD location in the dose range, denoted as $d^*$, 2) relative MTD location in the equivalence interval, $\lambda_1$, 3) distance from the lower bound of the equivalence interval to the next lower dose below MTD, $\rho_1$ , and 4) distance from the upper bound of the equivalence interval to the next higher dose above MTD, $\rho_2$. These factors will be thoroughly investigated in the simulation study later. \\

\vskip 0.1in
\noindent
Under $H_0$, we consider the following three cases for the sampling prior $\pi_0^{(s)}(\bp \mid H_0)$: 

\begin{itemize}
	\item[(Case i)] order statistics from the uniform distribution $\mathcal{U}(0, p_T - \epsilon_1)$, i.e., $\bp=(p_{(1)},p_{(2)},\ldots,p_{(D)})$, $p_d \sim \mathcal{U}(0,p_T-\epsilon_1), d=1,2,\ldots,D$;
	\item[(Case ii)] uniform distribution with monotonicity across doses, \\ i.e., $p_1 \sim \mathcal{U}(0,p_T-\epsilon_1), p_2 \sim \mathcal{U}(p_1,p_T-\epsilon_1), \cdots, p_D \sim \mathcal{U}(p_{D-1},p_T-\epsilon_1)$;
	\item[(Case iii)] point mass at $\bp=(p_T-\epsilon_1,\ldots,p_T-\epsilon_1)$, i.e., all doses have the same toxicity probability $(p_T-\epsilon_1)$ (the worst case for sample size calculation).
\end{itemize}

\begin{figure}
	\centering \includegraphics[width=0.85\textwidth]{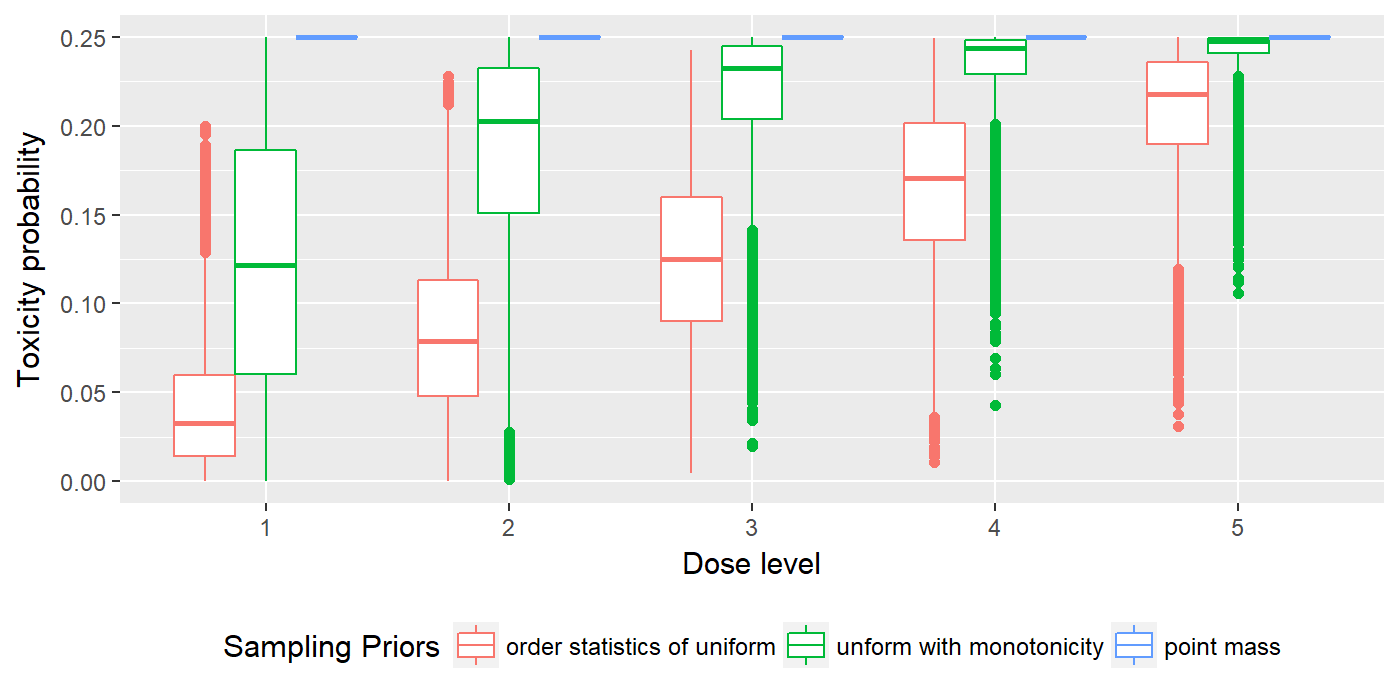}
	\caption{Three specific cases of the sampling priors under $H_0$ for $p_T=0.3$, $\epsilon_1=0.05$ and $D=5$.}
	\label{fig:sp0}
\end{figure}

\vskip 0.1in
\noindent
See Figure \ref{fig:sp0} for an illustration. Note that for all three cases above, the toxicity probabilities are restricted in the subspace given by the null hypothesis $H_0$, which is $\{p_d: p_d -p_T \le \epsilon_1\}$, i.e., all dose levels are lower than $(p_T - \epsilon_1)$. In the case where all dose levels are higher than $p_T$, trials are likely to be terminated early by the safety rules adopted in many designs, such as the mTPI-2 design \citep{guo2017bayesian}. For example, if $Pr(p_1 > p_T \mid data) > 0.95$, trial will be terminated early before reaching the sample size. Therefore, this case is not considered in specifying $\pi_0^{(s)}$. 

\subsection{Algorithm to Determine Sample Size} \label{Sec:scheme}

We develop Algorithm 1 to determine the optimal sample size $n^\star$ in Box \eqref{box} based on a numerical procedure. Given the target toxicity rate $p_T$, effect size $(\epsilon_1 + \epsilon_2)$, Type I error rate $\alpha$, Type II error rate $\beta$, and an upper bound for the maximum sample size $n_{U}$, the algorithm is initialized at $n_U$ and then seeks an optimal sample size $n^\star \le n_U$ iteratively. During the $j$-th iteration (assuming the current sample size is $n_j$), the algorithm takes two steps to identify the next proposed sample size $n_{j+1}$. In the first step, $B$ trials are simulated under the null hypothesis $H_0$ and the cut-off value $BF_0$ is calculated to control the prespecified type I error rate $\alpha$. Specifically, denote $BF_{0,b} = BF_0(\bY_b^{(n_j)})$, where $Y_n^{(n_j)}$ is the data of the $b$-th simulated trial under sample size $n_j$. Then $BF_0$ is chosen to be the $\alpha$-th quantile of $p(BF(\bY^{(n_j)})|H_0)$, the prior predictive distribution of $BF(\bY^{(n_j)})$ under $H_0$, denoted as $BF_{0, \alpha}$. Numerically, $BF_{0,\alpha} = BF_{0, (\floor{ B \cdot \alpha})}$, where $(\cdot)$ is the order statistic and $\floor{\cdot}$ is the floor function. In words, $BF_{0, \alpha}$ is the $\alpha$-th percentile of $\{BF_{0,b}; b=1, ..., B\}$. In the second step, trial data are simulated under the alternative hypothesis $H_1$ and the cut-off value $BF_0$ calculated from step 1 is used to calculate the power $\xi_1(n_j) = Pr(B(\bY^{(n_j)}) < BF_{0, \alpha} \mid H_1)$. Here $\xi_1(n_j)$ is approximated by simulating $C$ trials under $H_1$. If the power is above the prespecified threshold $(1-\beta)$ (i.e., the current sample size $n_j$ is too large), the algorithm updates the sample size $n_{j+1}$ to be the smallest integer that is greater than $\frac{n_j + n_j^-}{2}$ (i.e., $\ceil{(n_j + n_j^-) /2}$), where $n_j^-$ is the more recent sample size below $n_j$; if the power is below $(1-\beta)$ (i.e., the current sample size $n_j$ is too small), the next optimal sample size $n_{j+1}$ is updated to be the smallest integer that is larger than $\frac{n_j^+ + n_j}{2}$ (i.e., $\ceil{(n_j^+ + n_j) /2}$), where $n_j^+$ is the more recent sample size above $n_j$. The algorithm converges when $|n_{j+1} - n_{j}| < \epsilon$ for a given threshold $\epsilon$ and the final sample size is determined to be the last $n_{j+1}$. \\

\begin{algorithm}
	\label{BSSD_algorithm}
	\caption{BaySize Algorithm for Prespecified $p_T$, $(\epsilon_1, \epsilon_2)$, $\alpha$ and $\beta$}
	\begin{algorithmic}[1]
		\scriptsize
		\STATE Step 0: Initialization
			\STATE \quad Initialize the sample size at $n_1 = n_U$, where $n_U$ is the prespecified upper bound 
			\vskip 0.1in
		\STATE for $n_j < n_U$, $j>1$, do:
        \vskip 0.1in
		    \STATE \quad Step I: Type I error ($\alpha$) calibration
		        \STATE \quad  \quad for b = 1 to B (recommended B=1000)
		        \STATE \quad \quad  \quad a: Generate $\bp_b \sim \pi_0^{(s)}(\bp \mid H_0)$
		        \STATE \quad \quad  \quad b: Generate $\bY_b^{(n_j)} \sim f(\bY \mid \bp_b)$
		        \STATE \quad \quad  \quad c: Calculate $BF_{0,b} = BF_0 (\bY_b^{(n_j)})$
		        \STATE \quad  \quad Denote $S_0 = \{BF_{0,1}, ..., BF_{0,B}\}$ and $BF_{0, \alpha} = BF_{0, (\floor{B \cdot \alpha})}$ be the $\alpha$th percentile of $S_0$. Then $\hat{\xi}_0(n_j) = \frac{1}{B}\sum_{b=1}^B I_{\{BF_{0,b} \le BF_{0, \alpha}\}} \le \alpha$
		    \vskip 0.1in
	  \STATE  \quad Step II: Power $(1-\beta)$ verification
	       \STATE \quad  \quad for c = 1 to C (recommended C=1000)
	       \STATE \quad \quad  \quad a: Generate $\bp_c \sim \pi_1^{(s)}(\bp \mid H_1)$
	       \STATE \quad \quad  \quad b: Generate $\bY_c^{(n_j)} \sim f(\bY \mid \bp_c)$
	       \STATE \quad \quad  \quad c: Calculate $BF_{1,c} = BF_0 (\bY_c^{(n_j)})$
	       \STATE \quad  \quad Denote $S_1 = \{BF_{0,1}, ..., BF_{0,C}\}$. Then $\hat{\xi}_1(n_j) = \frac{1}{C}\sum_{c=1}^C I_{\{BF_{0,c} \le BF_{0, \alpha}\}}$
	       \STATE \quad  \quad if $\hat{\xi}_1 (n_j) > 1-\beta$,
	       \STATE \quad \quad  \quad $n_{j+1} =\ceil{(n_j + n_j^-) / 2}$, where $n_j^-$ is the most recent sample size below $n_j$;
	       \STATE \quad  \quad else if $\hat{\xi}_1 (n_j) \le 1-\beta$,
	       \STATE \quad \quad  \quad $n_{j+1} = \ceil{(n_j^+ + n_j)/2}$, where $n_j^+$ is the most recent sample size above $n_j$;
	       \vskip 0.1in
	\STATE 	 \quad Step III: repeat Step I and II until $|n_{j+1} - n_{j} | < \epsilon$, where $\epsilon$ is a prespecified threshold. 
	\vskip 0.1in
	\STATE Output: the final $n_{j+1}$
	\end{algorithmic}
\end{algorithm}

\section{Look-Up Table} \label{Sec:application}
For practical trials, the sample size determination is often conducted before a trial begins and as a result the toxicity probabilities $\bp_1^*$ under $H_1$ is unknown. Table \ref{tab:effectsize} calculates the power for each candidate sample size given a range of effect sizes and Type I error rates for $p_T=0.3$. For instance, when the half effect size $\epsilon_1 = \epsilon_2=0.1$, Type I error rate is controlled at 30\% and power no less than 80\%, the minimum sample size needed is $60$. Here the simulation-based searching algorithm for the BaySize framework adopts five different scenarios for $\bp_1^*$ where in each scenario, the true MTD locates at a different dose level. In Table \ref{tab:effectsize0.2} of Appendix \ref{Sec:otherapp}, we provide the sample size look-up table for $p_T=0.2$.  \\

\begin{table}[H]
	\centering
	\caption{For each combination of the half effect size $\epsilon_1=\epsilon_2$, Type I error rate $\alpha$ and candidate sample size, the range of the power $(1-\beta)$ given $p_T=0.3$ is calculated under five different scenarios of $\bp_1^*$, where for each scenario, the true MTD locates at a different dose level as reflected in Figure \ref{fig:full_power}.} \label{tab:effectsize}
	\begin{tabular}{c c c c c c c}
		\hline \hline
		\multirow{2}{*}{Half effect size}&\multirow{2}{*}{Type I error rate}&\multicolumn{5}{c}{Power min$\sim$max}\\ \cline{3-7}
		&&n=30&n=45&n=60&n=75&n=90\\
		\hline 
		0.05&0.05&0.11$\sim$0.20&0.09$\sim$0.17&0.09$\sim$0.18&0.10$\sim$0.19&0.12$\sim$0.21\\
		&0.1&0.18$\sim$0.25&0.18$\sim$0.25&0.20$\sim$0.29&0.26$\sim$0.35&0.30$\sim$0.40\\
		&0.2&0.34$\sim$0.45&0.38$\sim$0.45&0.47$\sim$0.53&0.53$\sim$0.57&0.60$\sim$0.63\\
		&0.3&0.51$\sim$0.62&0.58$\sim$0.61&0.68$\sim$0.70&0.71$\sim$0.75&0.73$\sim$0.79\\
		&0.4&0.63$\sim$0.68&0.71$\sim$0.77&0.76$\sim$0.82&0.79$\sim$0.85&0.79$\sim$0.87\\
		&0.5&0.74$\sim$0.82&0.79$\sim$0.85&0.81$\sim$0.88&0.84$\sim$0.91&0.86$\sim$0.92\\
		\hline
		0.1&0.05&0.09$\sim$0.31&0.18$\sim$0.36&0.24$\sim$0.42&0.33$\sim$0.46&0.42$\sim$0.52\\
		&0.1&0.23$\sim$0.49&0.37$\sim$0.53&0.43$\sim$0.57&0.52$\sim$0.60&0.62$\sim$0.67\\
		&0.2&0.46$\sim$0.59&0.59$\sim$0.69&0.69$\sim$0.72&0.75$\sim$0.77&0.78$\sim$0.80\\
		&0.3&0.61$\sim$0.71&0.73$\sim$0.77&0.80$\sim$0.85&0.81$\sim$0.88&0.85$\sim$0.92\\
		&0.4&0.71$\sim$0.79&0.84$\sim$0.90&0.85$\sim$0.92&0.87$\sim$0.94&0.88$\sim$0.95\\
		&0.5&0.82$\sim$0.91&0.86$\sim$0.93&0.89$\sim$0.96&0.90$\sim$0.97&0.90$\sim$0.97\\
		\hline
		0.15&0.05&0.24$\sim$0.59&0.40$\sim$0.62&0.51$\sim$0.67&0.60$\sim$0.71&0.67$\sim$0.75\\
		&0.1&0.33$\sim$0.62&0.48$\sim$0.69&0.68$\sim$0.77&0.71$\sim$0.78&0.80$\sim$0.82\\
		&0.2&0.58$\sim$0.77&0.73$\sim$0.80&0.86$\sim$0.90&0.84$\sim$0.86&0.87$\sim$0.90\\
		&0.3&0.73$\sim$0.85&0.86$\sim$0.90&0.86$\sim$0.90&0.88$\sim$0.94&0.88$\sim$0.95\\
		&0.4&0.75$\sim$0.85&0.88$\sim$0.95&0.90$\sim$0.96&0.90$\sim$0.97&0.90$\sim$0.97\\
		&0.5&0.87$\sim$0.92&0.92$\sim$0.98&0.96$\sim$0.99&0.96$\sim$0.99&0.96$\sim$0.99\\
		\hline
		0.2&0.05&0.37$\sim$0.64&0.62$\sim$0.72&0.77$\sim$0.81&0.81$\sim$0.86&0.83$\sim$0.87\\
		&0.1&0.67$\sim$0.79&0.80$\sim$0.83&0.84$\sim$0.86&0.85$\sim$0.88&0.87$\sim$0.91\\
		&0.2&0.73$\sim$0.82&0.88$\sim$0.92&0.88$\sim$0.93&0.90$\sim$0.95&0.90$\sim$0.95\\
		&0.3&0.90$\sim$0.96&0.90$\sim$0.96&0.91$\sim$0.97&0.91$\sim$0.97&0.91$\sim$0.98\\
		&0.4&0.90$\sim$0.96&0.96$\sim$0.99&0.96$\sim$0.99&0.96$\sim$0.99&0.97$\sim$0.99\\
		&0.5&0.91$\sim$0.98&0.96$\sim$0.99&0.96$\sim$0.99&0.97$\sim$0.99&0.97$\sim$0.99\\
		\hline\hline
	\end{tabular}
\end{table}

\section{Simulation Studies} \label{Sec:sim}

\subsection{Simulation Setup} 
\vskip 0.1in
\noindent
Without loss of generality, we consider trials with the target toxicity rate $p_T=0.3$ and the number of dose levels $D=5$. Trials are simulated using the mTPI-2 design \citep{guo2017bayesian} with the first dose as the starting dose and a cohort size of 3. Type I error rate is controlled at various prespecified levels ranging between $0.05$ and $0.50$. Four values are considered for half effect size $\epsilon_1 = \epsilon_2 = (0.05, 0.10, 0.15, 0.20)$. For each effect size, five scenarios are considered for $\bp_1^*$ under $H_1$ where the true MTD corresponds to a different dose level. Specifically, for each scenario of $\bp_1^*$, the toxicity probability of the true MTD is set to be $p_T$, i.e., the true MTD lies exactly in the middle of the EI with $p_{d^*} = p_T$, $p_{d^*-1} = p_T - \epsilon_1$ and $p_{d^*+1} = p_T + \epsilon_2$, where $d^*$ is the true MTD. For the sampling prior $\pi_0^{(s)}(\bp \mid H_0)$ under $H_0$, we use the order statistics from the uniform distribution (Case (i) in Section \ref{Sec:sp}). Finally, vague fitting priors are used for both $H_0$ and $H_1$ with $c=0$ in equation \eqref{eq:beta1} and \eqref{eq:beta2}. Two types of simulations are conducted. First, sample size is calculated according to Algorithm 1 for a range of Type I error rate and power. A practical consideration in trial design is to set the maximum sample size to be the multiple of dose levels. Therefore, in the second type of simulations, a set of candidate sample sizes are specified and statistical power is calculated for each combination of Type I error rate and candidate sample size. We set up scenarios with varying factors, such as the MTD location in the EI ($\lambda_1$), and the distance from the lower / upper bound of the EI to the adjacent doses ($\rho_1$ and $\rho_2$), as shown in Figure \ref{fig:sp1}. Finally, sensitivity analyses are conducted to evaluate different specifications of the sampling prior $\pi_0^{(s)}(\bp \mid H_0)$ and the fitting prior $\pi_1^{(f)}(\bp \mid H_0)$ and $\pi_1^{(f)}(\bp \mid H_1)$, e.g., different specifications of dispersion parameter $c$ and mode vector $\bm{q^{1d}}$ and $\bm{q^{1d}}$ in equation \eqref{eq:beta1} and \eqref{eq:beta2}. \\

\subsection{Simulation Results}
\begin{figure}
	\centering
	\subfigure[Sample Size]{
		\label{fig:main_a}
		\includegraphics[width=0.4\textwidth]{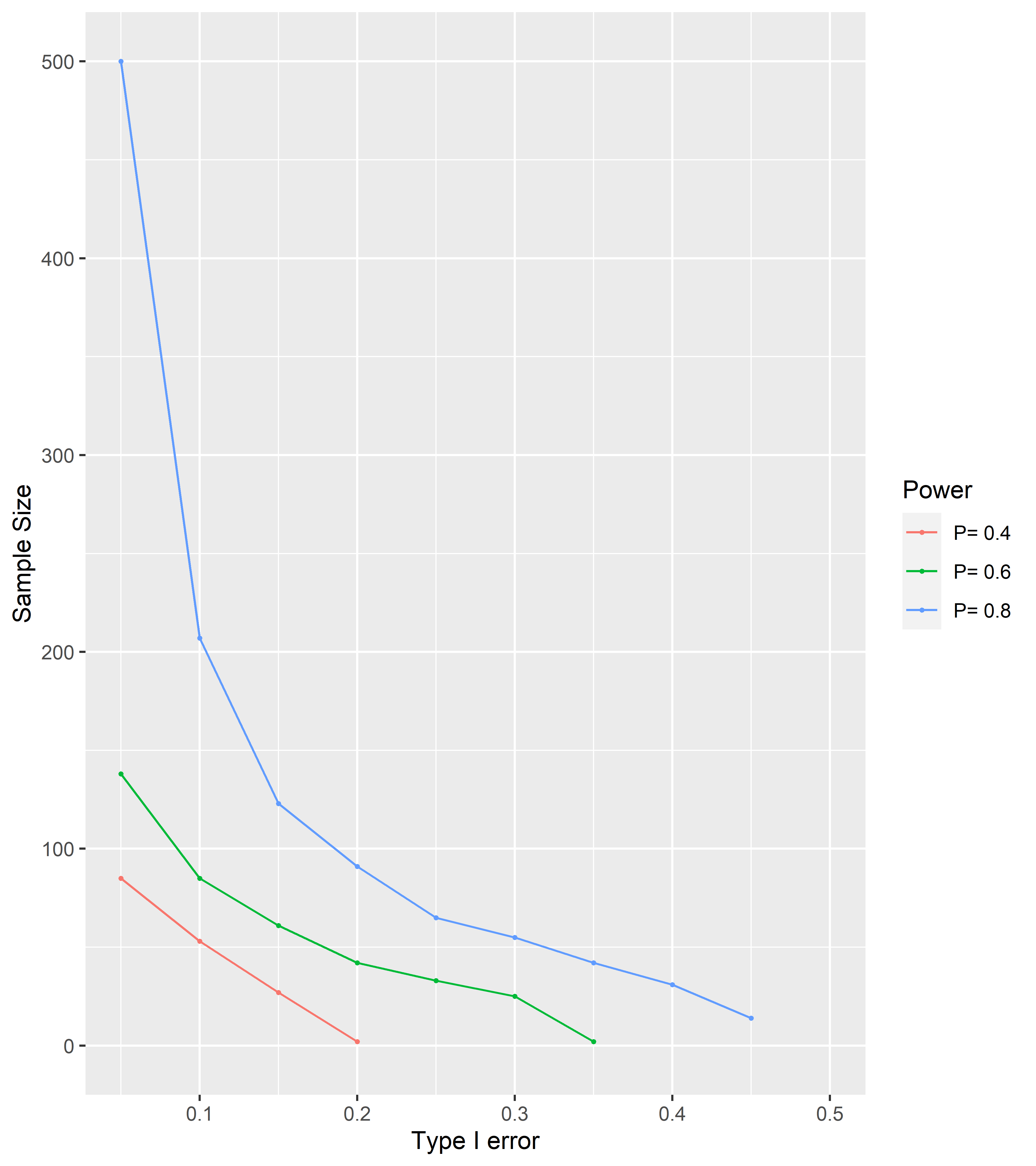}
	}
	\subfigure[Statistical power]{
		\label{fig:main_b}
		\includegraphics[width=0.4\textwidth]{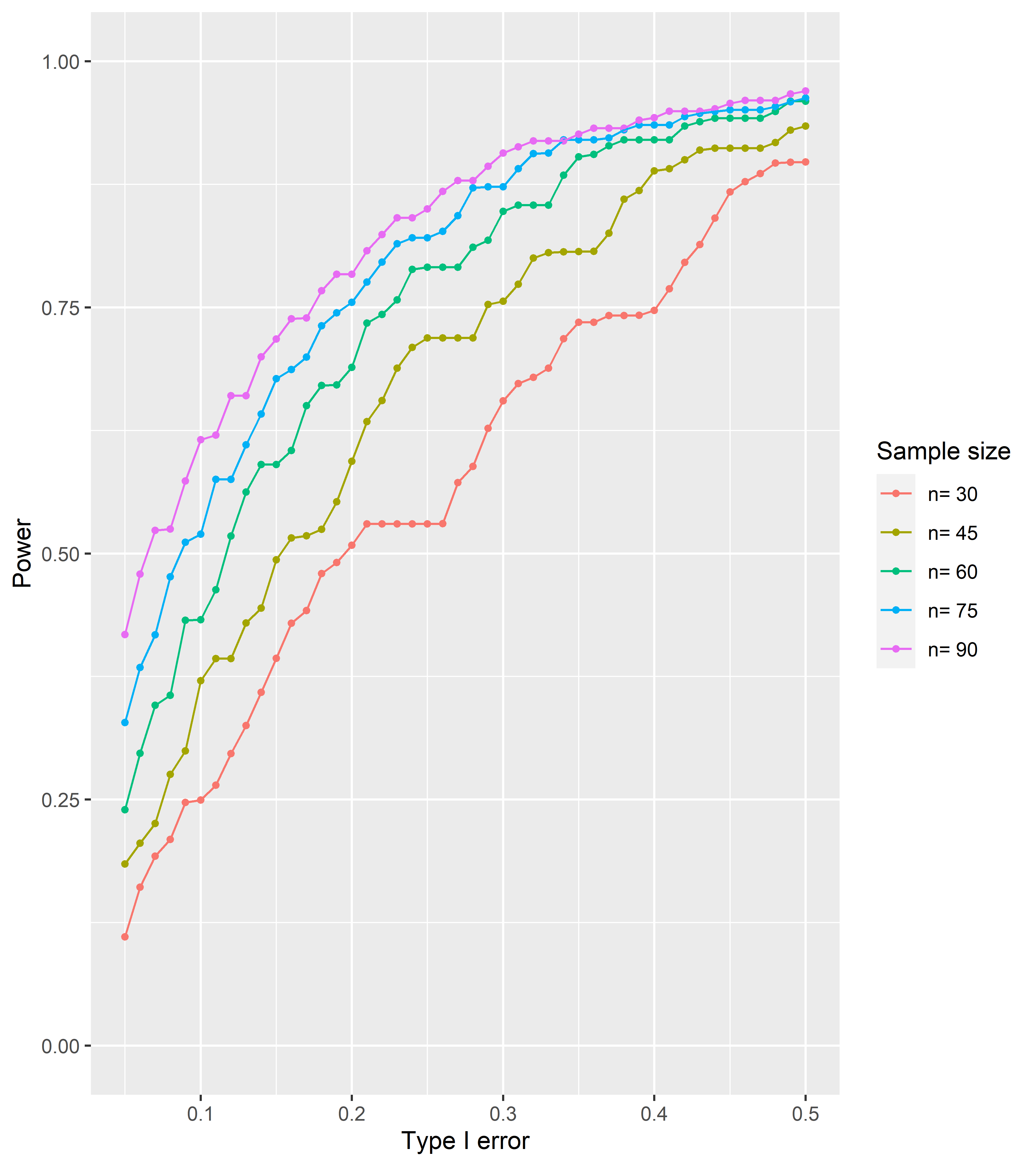}
	}	
	\caption[Short Title]%
	{Sample size and statistical power calculated according to BaySize given $p_T=0.3$, half effect size $\epsilon_1 = \epsilon_2=0.1$, $\bp_1^* = (0.1, 0.2, 0.3, 0.4, 0.5)$ and Type I error rate controlled between 5\% and 50\% . (a) \small Sample size when statistical power equals 40\%, 60\% and 80\%. (b) Statistical power when candidate sample size equals 30, 45, 60, 75 and 90..}
\end{figure}

Figure \ref{fig:main_a} and \ref{fig:main_b} demonstrate the sample size and statistical power calculated by BaySize when power equals 40\%, 60\%, 80\%, and candidate sample size equals 30, 45, 60, 75 and 90, for $p_T=0.3$, half effect size $\epsilon_1 = \epsilon_2=0.1$, $\bp_1^* = (0.1, 0.2, 0.3, 0.4, 0.5)$ and Type I error rate controlled between 5\% and 50\%

\vskip 0.1in
\noindent
As expected, for fixed $p_T$, effect size $(\epsilon_1 + \epsilon_2)$, $\bp_1^*$ and Type I error rate $\alpha$, sample size increases with power, shown in Figure \ref{fig:main_a}. For example, when Type I error rate is set at $\alpha=0.15$, one needs to enroll 29, 65 and 123 subjects to achieve 40\%, 60\% and 80\% statistical power, respectively. Taking a different angle, for fixed $p_T$, effect size $(\epsilon_1 + \epsilon_2)$, $\bp_1^*$ and Type I error rate $\alpha$, power increases with sample size, as is reflected in Figure \ref{fig:main_b}. For example, when the Type I error rate $\alpha = 0.3$, statistical power increases from 65.50\% to 75.64\%, 84.77\%, 87.25\%, 90.70\% as the sample size increases from 30 to 45, 60, 75, 90. An important observation from these results is that the gain of statistical power diminishes as sample size increases. In Figure \ref{fig:main_b}, statistical power increases by 10.14\%, 9.13\%, 2.48\% and 3.45\% when the sample size increases every 15 units from 30 to 45, 60, 75 and 90. Full simulation results for the sample size and power calculation for each combination of the effect size and $\bp_1^*$ is shown in Figure \ref{fig:full_ss} and \ref{fig:full_power} of Appendix \ref{sec:full_ss}. \\

\vskip 0.1in
\noindent
Next, we consider the effect of relative MTD location on the sample size and statistical power. As shown in Figure \ref{fig:full_ss} and \ref{fig:full_power}, the lower the true MTD location, the easier it is to be identified, and thus the smaller the sample size with fixed power, or the higher the power with fixed sample size. This is expected since dose finding usually starts at the lowest dose and the MTD is more difficult to reach in dose finding trials when it is located at a higher dose level. It should also be noticed that given an effect size $(\epsilon_1 + \epsilon_2)$ and sample size $n$, the gain of power depends on $\bp_1^*$ as well. For instance, consider $\epsilon_1 = \epsilon_2 = 0.1$ (the second row in Figure \ref{fig:full_power}) and sample size $n=30$ (the red line in Figure \ref{fig:full_power}), power increases by 52.69\% (from 30.56\% to 83.25\%) for $\bp_1^* = (0.3, 0.4, 0.5, 0.6, 0.7)$, while 69.6\% (from 12.28\% to 81.88\%) for $\bp_1^* = (0.01, 0.05, 0.1, 0.2, 0.3)$. There are possibly two explanations. First, for $\bp_1^*=(0.3, 0.4, 0.5, 0.6, 0.7)$ where the first dose is the true MTD and all doses above are overly toxic, the algorithm in mTPI-2 design basically allocates the most patients to the first two doses and almost no patients to the last three doses. While for $\bp_1^* = (0.01, 0.05, 0.1, 0.2, 0.3)$ where all doses are safe and the last dose is considered as the true MTD, patients are allocated to all five doses and inference about the toxicity probability $p_d$ is done at all dose levels, which facilitates the fast increasing of the power. Secondly, scenarios with $\bp_1^* = (0.3, 0.4, 0.5, 0.6, 0.7)$ are more likely to be terminated by the safety rules in many designs such as mTPI-2, and as a result, the average number of patients enrolled and treated is fewer than the maximum sample sizes prespecified. This is particularly obvious from the sub-figure (row 2, column 1) in Figure \ref{fig:full_power} that the power barely increases from $n=75$ to $n=90$. \\

\vskip 0.1in
\noindent
Finally, if we compare each column in Figure \ref{fig:full_ss} and \ref{fig:full_power} with the same $\bp_1^*$ and different effect sizes $(\epsilon_1 + \epsilon_2)$, it is easy to find that given a fixed power, fewer patients are needed for larger effect size, and given a fixed candidate sample size, higher power is achieved for larger effect size. This indicates that MTD is easier to be identified for larger effect sizes, possibly due to the reason that given $p_T=0.3$, it is easier to distinguish the toxicity probability between $0.3$ and $0.15$ than between $0.3$ and $0.25$. In addition, for large effect size such as $\epsilon_1 = \epsilon_2=0.2$ (the last row in Figure \ref{fig:full_ss} and \ref{fig:full_power}), the power quickly reaches its plateau (close to 1) for all candidate sample sizes when $\alpha$ increases from $0.05$ to $0.5$. In other words, the power gain in the context of large effect sizes is limited compared with small effect sizes. \\

\vskip 0.1in
\noindent
Simulation results for $p_T=0.2$ are provided in Appendix \ref{otherpt} and similar patterns are observed. In addition, various factors that could potentially affect the sample size determination (and power calculation) are thoroughly investigated and results are provided in Appendix \ref{otherfactor}. In summary, \ref{fig:lambda-1} and \ref{fig:lambda-2} indicate that fewer patients are needed if the true MTD locates in the middle of the equivalence interval, while Figures \ref{fig:rho1-1}, \ref{fig:rho1-2}, \ref{fig:rho2-1} and \ref{fig:rho2-2} indicate that distances from the lower / upper bounds of the equivalence interval to the adjacent doses below / above the MTD has little impact on the sample size determination. \\

\subsection{Sensitivity Analysis}
The effects of different sampling priors $\pi_0^{(s)}(\bp \mid H_0)$, dispersion parameters $c$ and mode vectors ($\bm{q^{1d}}$ and $\bm{q^{0d}}$) of the fitting priors are investigated through a series of sensitivity analyses. Results are provided in Appendix \ref{sensitivity}. In brief, Figures \ref{fig:sampling-1} indicates that the specification of $\pi_0^{(s)}(\bp \mid H_0)$ has a large impact on the power calculation and thus on sample size determination. Given $p_T=0.3$ and $\epsilon_1 = \epsilon_2 = 0.1$, for a fixed $\bp_1^*$ and $\alpha$, order statistics from the uniform distribution (Case i in Section \ref{Sec:sp}) always has the largest power for a given sample size $n$, and thus the smallest sample size given a prespecified power. This is because the doses generated from the order statistics sampling prior are more dispersed and further away from $p_T$, and as a result, it is easier to accept $H_0$ under the null. With a large calibrated $BF_0$ in step 1 of Algorithm 1, one would get a larger power for each $n$ (and small $n$ for each prespecified power level). Figure \ref{fig:cmode-1} and \ref{fig:cmode-2} indicate that given $p_T = 0.3$ and $\epsilon_1 = \epsilon_2 = 0.1$, the effect of fitting prior is overall small. \\

\section{Discussion}\label{Sec:dis}
In this paper, we propose a general Bayesian framework for sample size
determination in Phase I dose-finding trials, through formal Bayesian
  decision making under two composite hypotheses.   The
performance of BaySize was evaluated via extensive simulation studies
and results show that for a prespecified effect size and fitting /
sampling priors, BaySize is able to control Type I error rate while
achieving enough power.   Note that the ``power'' in BaySize does not
imply the correct selection of the true MTD, an objective of
dose-finding designs rather than sample size calculation. What BaySize
is concerned about is to empower the dose-finding trial so that when
the true MTD is present, the pre-planned sample size will allow
investigators to identify it with a sound statistical model and a design, such as
the fitting prior and the mTPI-2 design proposed in this paper.   \\

\vskip 0.1in
\noindent
The BaySize approach can account for parameter uncertainties through
prior specifications of the toxicity profiles. Specifically, for
fitting priors in the Bayes Factor thresholding, the model space is
augmented and divided into multiple sub-spaces in order to facilitate
the prior construction; while for sampling priors in trial data
generation, historical trial data or information about relevant trials
can be incorporated for better specifications of the toxicity
profiles. Besides, BaySize determines the minimum required sample size
according to the prespecified Type I error rate and power. Although
the sample size determination relies on a simulation-based searching
algorithm and does not have a closed form solution, the fast computing
speed of modern computers makes this method easy to implement.
\\

\vskip 0.1in
\noindent
Although the development of BaySize is illustrated using the mTPI-2 design throughout this paper, it can be easily extended to other designs as well, such as CRM and BOIN. In making inferences about the toxicity profiles and conducting hypothesis testing, BaySize collects the number of DLTs and total number of patients allocated at each dose. While trial data are simulated under a specific design, the simulation process is generally separated from the MTD identification and thus sample size determination. Therefore, BaySize is generally applicable as long as the design algorithms can output the toxicity outcomes given a prespecified maximum sample size. Furthermore, BaySize can be extended to time-to-event designs, such as TITE-CRM, TITE-BOIN and POD-TPI, where patients accrual can be continued with late-on-set toxicities via time to event modeling \citep{TITE_CRM, TITE_BOIN, POD_TPI}. Finally, BaySize can be extended to also include efficacy outcomes, which is important for dosing scenarios with non-monotonic dose-response relationship. \\

\newpage
\bibliographystyle{unsrtnat}
\bibliography{bib-bssd}

\newpage
\begin{appendices}
\appendixpage
\renewcommand\thefigure{\thesection.\arabic{figure}}   
\renewcommand\thetable{\thesection.\arabic{table}}
\renewcommand{\theequation}{\thesection.\arabic{equation}}
\setcounter{figure}{0}  
\setcounter{table}{0}
\setcounter{equation}{0}

\section{Full Simulation Results} \label{sec:full_ss}
\begin{figure}
	\centering \includegraphics[width=0.9\textwidth]{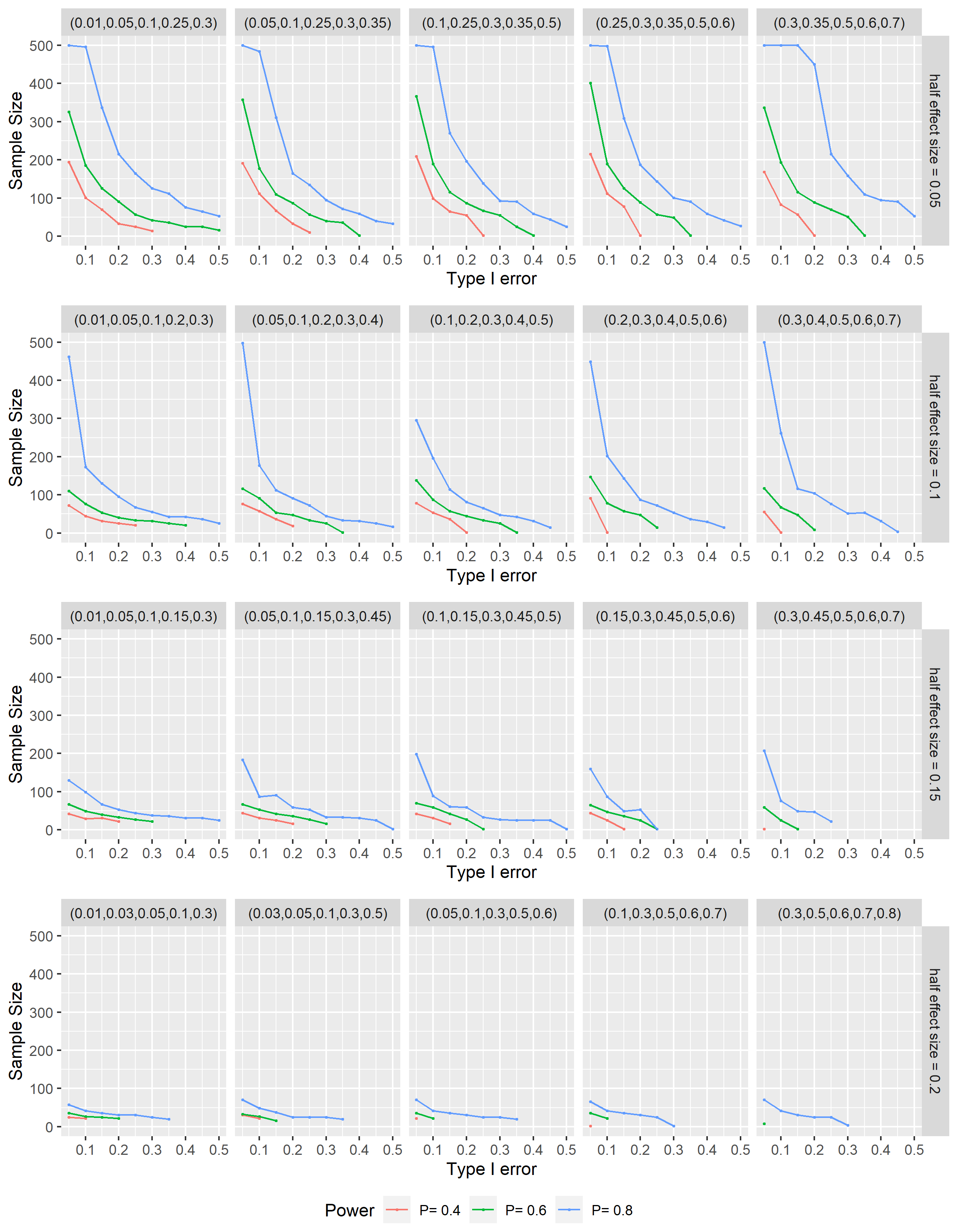}
	\caption{Sample size calculated for each combination of effect size, $\bp_1^*$ and candidate sample size when the Type I error rate ranges between $5\%$ to $50\%$. Each row indexes a half effect size among $\epsilon_1 = \epsilon_2 = (0.05, 0.10, 0.15, 0.20)$ and each column indexes one of five scenarios for $\bp_1^*$, in which the true MTD locates at a different dose level. In each sub-figure, each color indexes a power among $(0.4, 0.6, 0.8)$.}	
	\label{fig:full_ss}
\end{figure}

Full simulation results comparing the sample size calculation for each combination of half effect size and $\bp_1^*$ are summarized in Figure \ref{fig:full_ss}. \\

\begin{figure}
	\centering \includegraphics[width=0.9\textwidth]{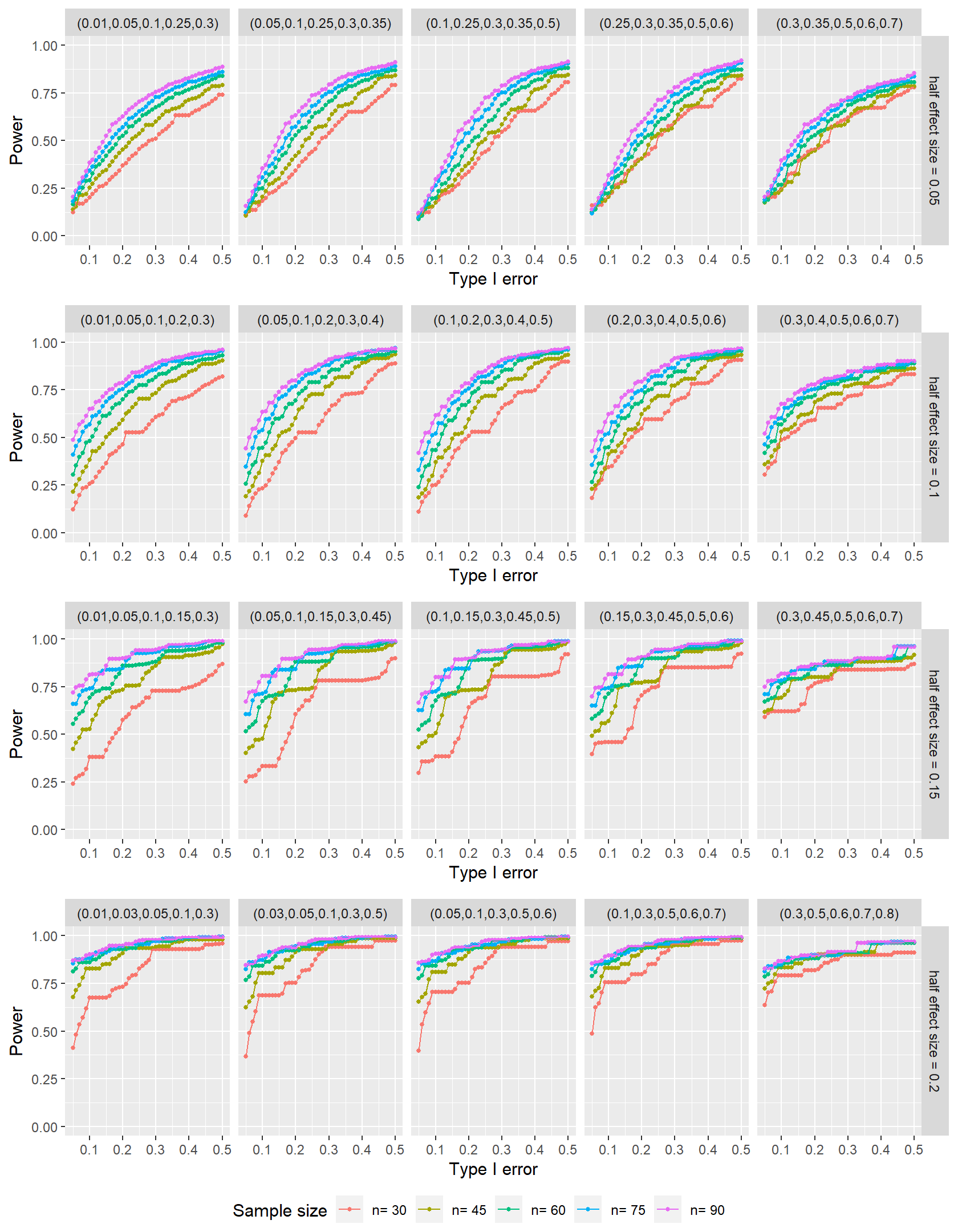}
	\caption{Power calculated for each combination of effect size $\epsilon_1 = \epsilon_2$, $\bp_1^*$ and candidate sample size when the Type I error rate range between $5\%$ to $50\%$. Each row indexes a half effect size among $\epsilon_1 = \epsilon_2 = (0.05, 0.10, 0.15, 0.20)$ and each column indexes one of five scenarios for $\bp_1^*$, in which the true MTD locates at a different dose level. In each sub-figure, each colors indexes a candidate sample size among $n=(30,45,60,75,90)$.}
	\label{fig:full_power}
\end{figure}

\vskip 0.1in
\noindent
Full simulation results comparing the statistical power for each combination of half effect size and $\bp_1^*$ are summarized in Figure \ref{fig:full_power}. \\

\section{Simulation for $p_T=0.2$} \label{otherpt}
\begin{figure}
	\centering \includegraphics[width=0.9\textwidth]{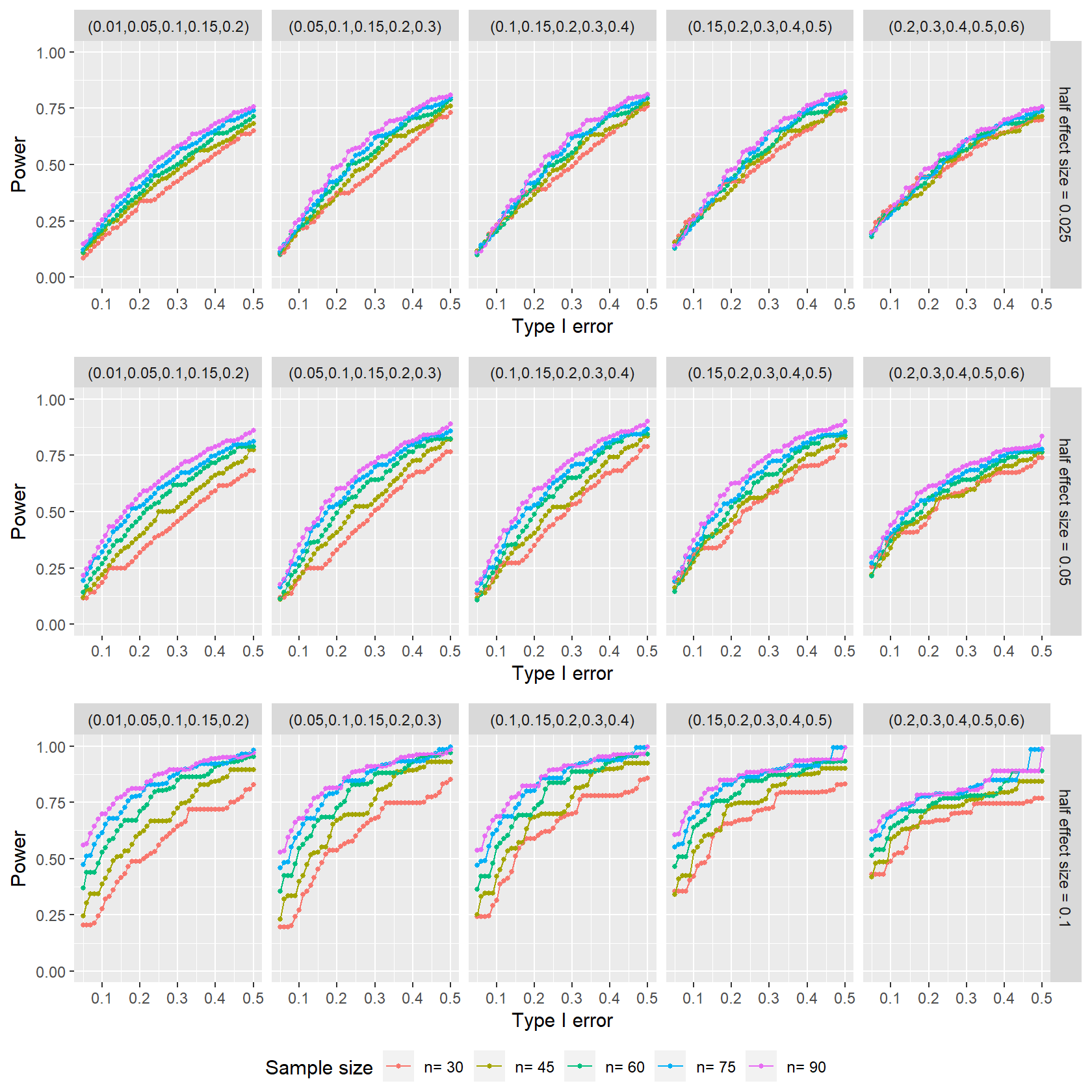}
		\caption{Power calculated for each combination of effect size, $\bp_1^*$ and candidate sample size when the Type I error rate ranges between $5\%$ to $50\%$. Each row indexes a half effect size among $\epsilon_1 = \epsilon_2 = (0.05, 0.10, 0.15, 0.20)$ and each column indexes one of five scenarios for $\bp_1^*$, in which the true MTD locates at a different dose level. In each sub-figure, each color indexes a candidate sample size among $n=(30,45,60,75,90)$.}
	\label{fig:effectsize_0.2}
\end{figure}

Simulation studies are conducted for $p_T=0.2$. Results are summarized in Figure \ref{fig:effectsize_0.2}.\\

\section{Other Factors that Could Impact the Sample Size Determination} \label{otherfactor}
\begin{figure}
	\centering \includegraphics[width=0.9\textwidth]{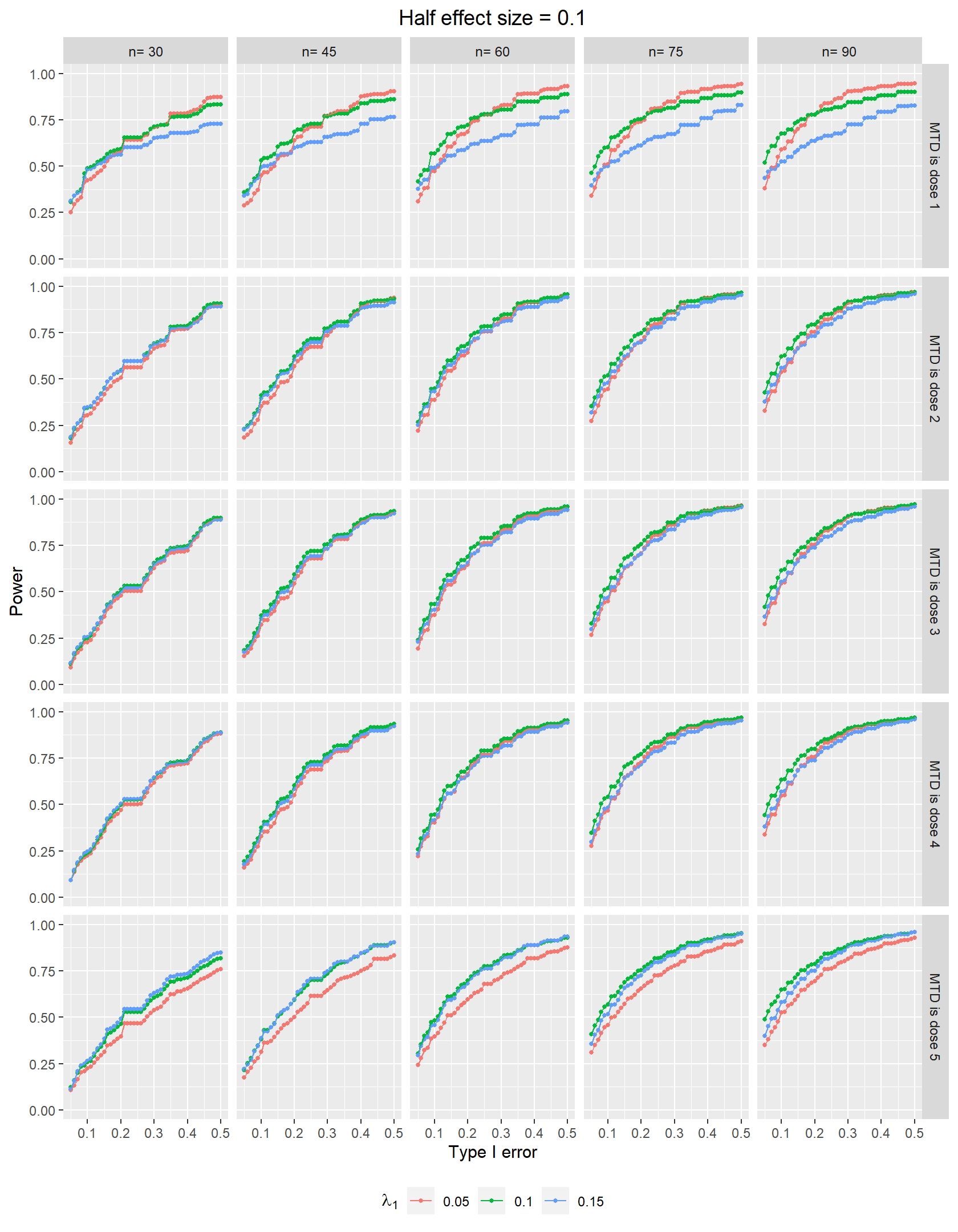}
	\caption{Power calculated for each combination of $\lambda_1$ and candidate sample size when the Type I error rate ranges between $5\%$ to $50\%$ and half effect size $\epsilon_1=\epsilon_2=0.1$. Each row indexes an alternative scenario for $\bp_1^*$ and each column indexes a candidate sample size. In each sub-figure, each color indexes a different value for $\lambda_1$.}
	\label{fig:lambda-1}
\end{figure}

\begin{figure}
	\centering \includegraphics[width=0.9\textwidth]{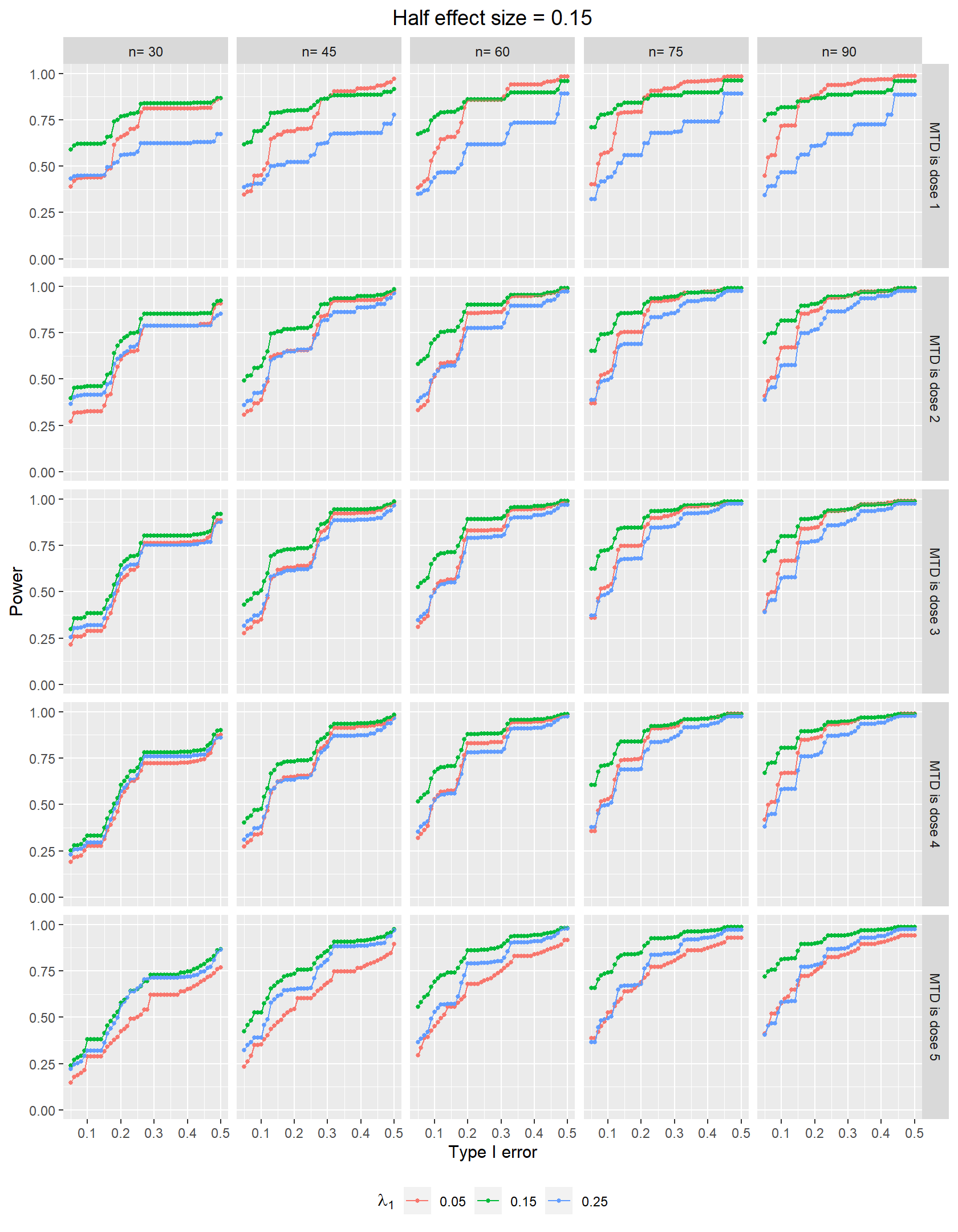}	
	\caption{Power calculated for each combination of $\lambda_1$ and candidate sample size when the Type I error rate ranges between $5\%$ to $50\%$ and half effect size $\epsilon_1=\epsilon_2=0.15$. Each row indexes an alternative scenario for $\bp_1^*$ and each column indexes a candidate sample size. In each sub-figure, each color indexes a different value for $\lambda_1$.}
	\label{fig:lambda-2}
\end{figure}

\begin{figure}
	\centering \includegraphics[width=0.9\textwidth]{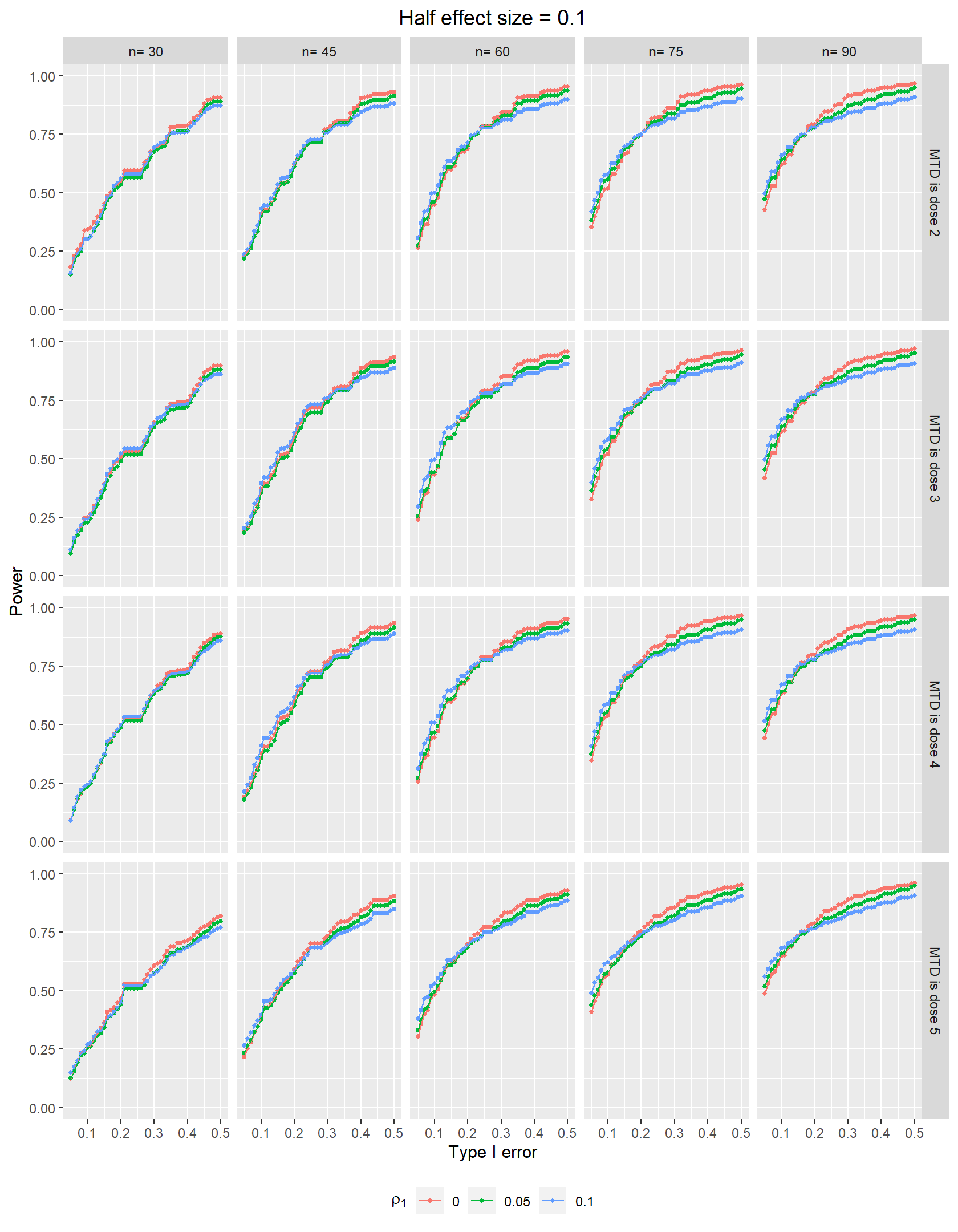}	
	\caption{Power calculated for each combination of $\rho_1$ and candidate sample size when the Type I error rate ranges between $5\%$ to $50\%$ and half effect size $\epsilon_1=\epsilon_2=0.1$. Each row indexes an alternative scenario for $\bp_1^*$ and each column indexes a candidate sample size. In each sub-figure, each color indexes a different value for $\rho_1$.}
	\label{fig:rho1-1}
\end{figure}

\begin{figure}
	\centering \includegraphics[width=0.9\textwidth]{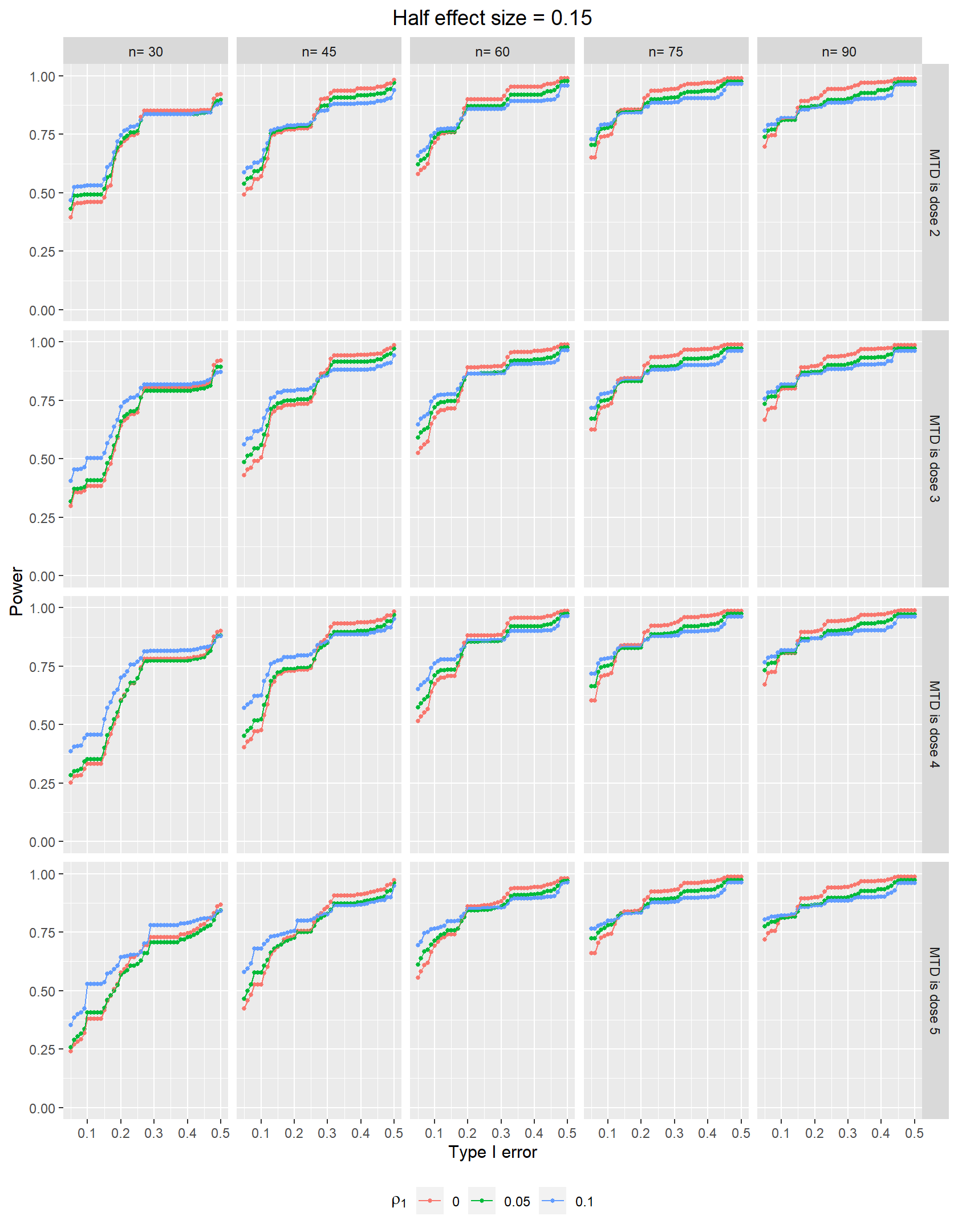}
	\caption{Power calculated for each combination of $\rho_1$ and candidate sample size when the Type I error rate ranges between $5\%$ to $50\%$ and half effect size $\epsilon_1=\epsilon_2=0.15$. Each row indexes an alternative scenario for $\bp_1^*$ and each column indexes a candidate sample size. In each sub-figure, each color indexes a different value for $\rho_1$.}
	\label{fig:rho1-2}
\end{figure}

\begin{figure}
	\centering \includegraphics[width=0.9\textwidth]{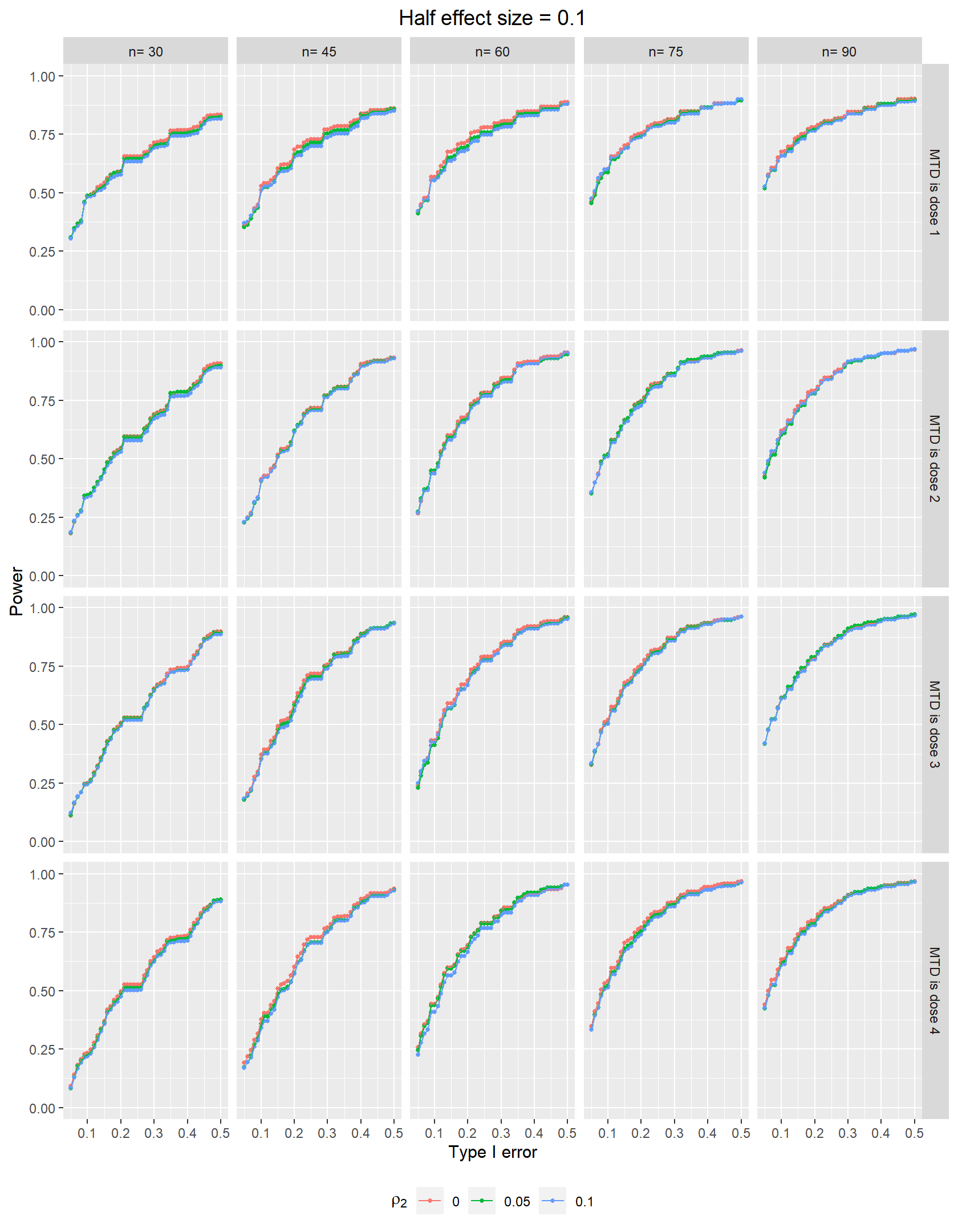}
	\caption{Power calculated for each combination of $\rho_2$ and candidate sample size when the Type I error rate ranges between $5\%$ to $50\%$ and half effect size $\epsilon_1=\epsilon_2=0.1$. Each row indexes an alternative scenario for $\bp_1^*$ and each column indexes a candidate sample size. In each sub-figure, each color indexes a different value for $\rho_2$.}	
	\label{fig:rho2-1}
\end{figure}

\begin{figure}
	\centering \includegraphics[width=0.9\textwidth]{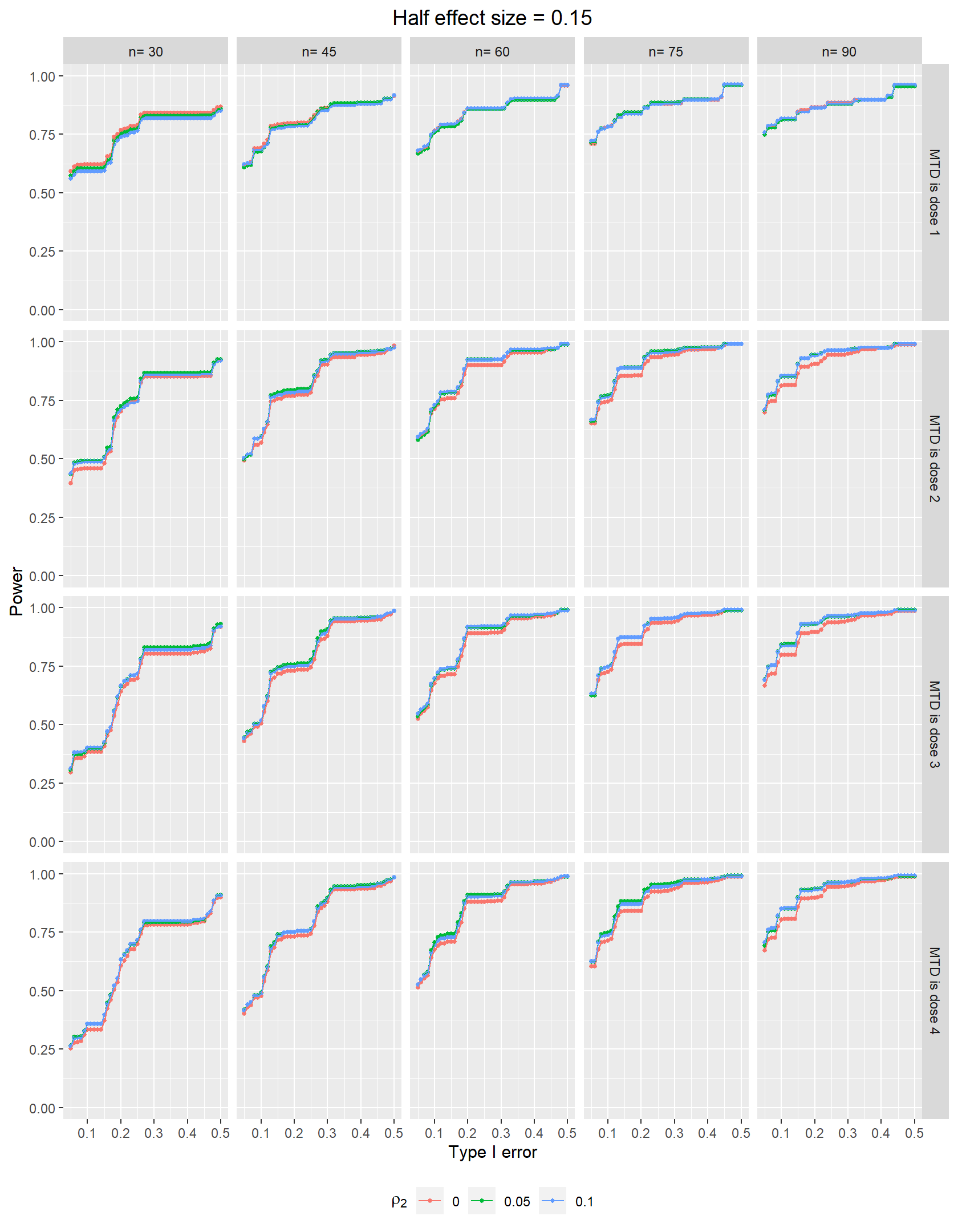}
	\caption{Power calculated for each combination of $\rho_2$ and candidate sample size when the Type I error rate ranges between $5\%$ to $50\%$ and half effect size $\epsilon_1=\epsilon_2=0.15$. Each row indexes an alternative scenario for $\bp_1^*$ and each column indexes a candidate sample size. In each sub-figure, each color indexes a different value for $\rho_2$.}	
	\label{fig:rho2-2}
\end{figure}

Simulation studies are conducted to investigate the effect of MTD location in the EI ($\lambda_1$), and the distances from the lower / upper boundary of the EI to the next dose below / above the MTD ($\rho_1$ and $\rho_2$). Trials are simulated under half effect size $\epsilon_1=\epsilon_2=0.1$ and $\epsilon_1=\epsilon_2=0.15$, respectively. Results are shown in Figures \ref{fig:lambda-1}, \ref{fig:lambda-2}, \ref{fig:rho1-1}, \ref{fig:rho1-2}, \ref{fig:rho2-1} and \ref{fig:rho2-2}. \\

\vskip 0.1in
\noindent
Overall, $\rho_1$ and $\rho_2$ have limited impact on the statistical power and sample size determination, according to Figure \ref{fig:rho1-1}, \ref{fig:rho1-2}, \ref{fig:rho2-1} and \ref{fig:rho2-2}. Figures \ref{fig:lambda-1} and \ref{fig:lambda-2} indicate that, for scenarios where the first or the last dose is the true MTD, larger power (given the candidate sample size) or smaller sample size (given the prespecified power rate) is observed when the true MTD locates toward the middle of the EI, i.e., when $\lambda_1$ moves further away from 0. \\

\newpage
\section{Sensitivity analysis}
\subsection{A simple specification of $\bm{q^{ij}}$} \label{modes}
The vectors $\bm{q^{ij}}=(q^{ij}_1,\cdots,q^{ij}_D)$ used in the fitting priors in \eqref{eq:beta1} and \eqref{eq:beta2} are the modes of $D$ truncated Beta distributions under different augmented model $M_{ij}$, $i=0,1$, $j=0,1,\cdots,D$, if $i=0$ and $j=1,\cdots,D$, if $i=1$, which can be simply elicited as follows: 
\begin{equation}
q^{ij}_k=
\begin{cases}
a_1*(p_T-\epsilon_1), & \mbox{if~} i=1, k<j-1 \mbox{~or~} i=0, k<D-j\\
a_2*(p_T-\epsilon_1), & \mbox{if~} i=1, k=j-1 \mbox{~or~} i=0, k=D-j\\
p_T, & \mbox{if~} i=1,k=j \\
a_3*(p_T+\epsilon_2), & \mbox{if~} i=1, k=j+1 \mbox{~or~} i=0, k=D-j+1\\
a_4*(p_T+\epsilon_2), & \mbox{if~} i=1, k>j+1 \mbox{~or~} i=0, k>D-j+1
\end{cases}
\quad
\begin{array}{l@{}l}
k{}=1,2,\cdots,D; \\
j{}=1,\cdots,D, \mbox{~if~} i=1;\\
j{}=0,1,\cdots,D, \mbox{~if~} i=0, 
\end{array} \label{eq:mode}
\end{equation}
where $0<a_1,a_2<1$ and  $a_3,a_4>1$. Note that we use $p_T - \epsilon_1$ and $p_T + \epsilon_2$ in \eqref{eq:mode} to force each element of the mode vector to be located within each corresponding truncated interval. Table \ref{ex.mode} shows an example when $D=5$, $p_T=0.30$, $\epsilon_1=\epsilon_2=0.10$, $a_1=0.6$, $a_2=0.9$, $a_3=1.05$ and $a_4=1.2$. 
\begin{table}[H]
	\centering
	\caption{An example of modes $\bm{q^{ij}}$ in the fitting priors}\label{ex.mode}
	\begin{tabular}{cccccccccccc}
		\hline
		$M_{ij}$&$M_{11}$&$M_{12}$&$M_{13}$&$M_{14}$&$M_{15}$&$M_{00}$&$M_{01}$&$M_{02}$&$M_{03}$&$M_{04}$&$M_{05}$ \\
		\hline
		$q^{ij}_{k}$&$\bm{q^{11}_{\cdot}}$&$\bm{q^{12}_{\cdot}}$&$\bm{q^{13}_{\cdot}}$&$\bm{q^{14}_{\cdot}}$&$\bm{q^{15}_{\cdot}}$&$\bm{q^{00}_{\cdot}}$&$\bm{q^{01}_{\cdot}}$&$\bm{q^{02}_{\cdot}}$&$\bm{q^{03}_{\cdot}}$&$\bm{q^{04}_{\cdot}}$&$\bm{q^{05}_{\cdot}}$\\
		\hline
		$\bm{q^{\cdot\cdot}_{1}}$&0.30&0.18&0.12&0.12&0.12&0.12&0.12&0.12&0.12&0.18&0.42\\
		$\bm{q^{\cdot\cdot}_{2}}$&0.42&0.30&0.18&0.12&0.12&0.12&0.12&0.12&0.18&0.42&0.48\\
		$\bm{q^{\cdot\cdot}_{3}}$&0.48&0.42&0.30&0.18&0.12&0.12&0.12&0.18&0.42&0.48&0.48\\
		$\bm{q^{\cdot\cdot}_{4}}$&0.48&0.48&0.42&0.30&0.18&0.12&0.18&0.42&0.48&0.48&0.48\\
		$\bm{q^{\cdot\cdot}_{5}}$&0.48&0.48&0.48&0.42&0.30&0.18&0.42&0.48&0.48&0.48&0.48\\
		\hline
	\end{tabular}
\end{table}

\subsection{Simulation results} \label{sensitivity} 
\begin{figure}
	\centering \includegraphics[width=0.9\textwidth]{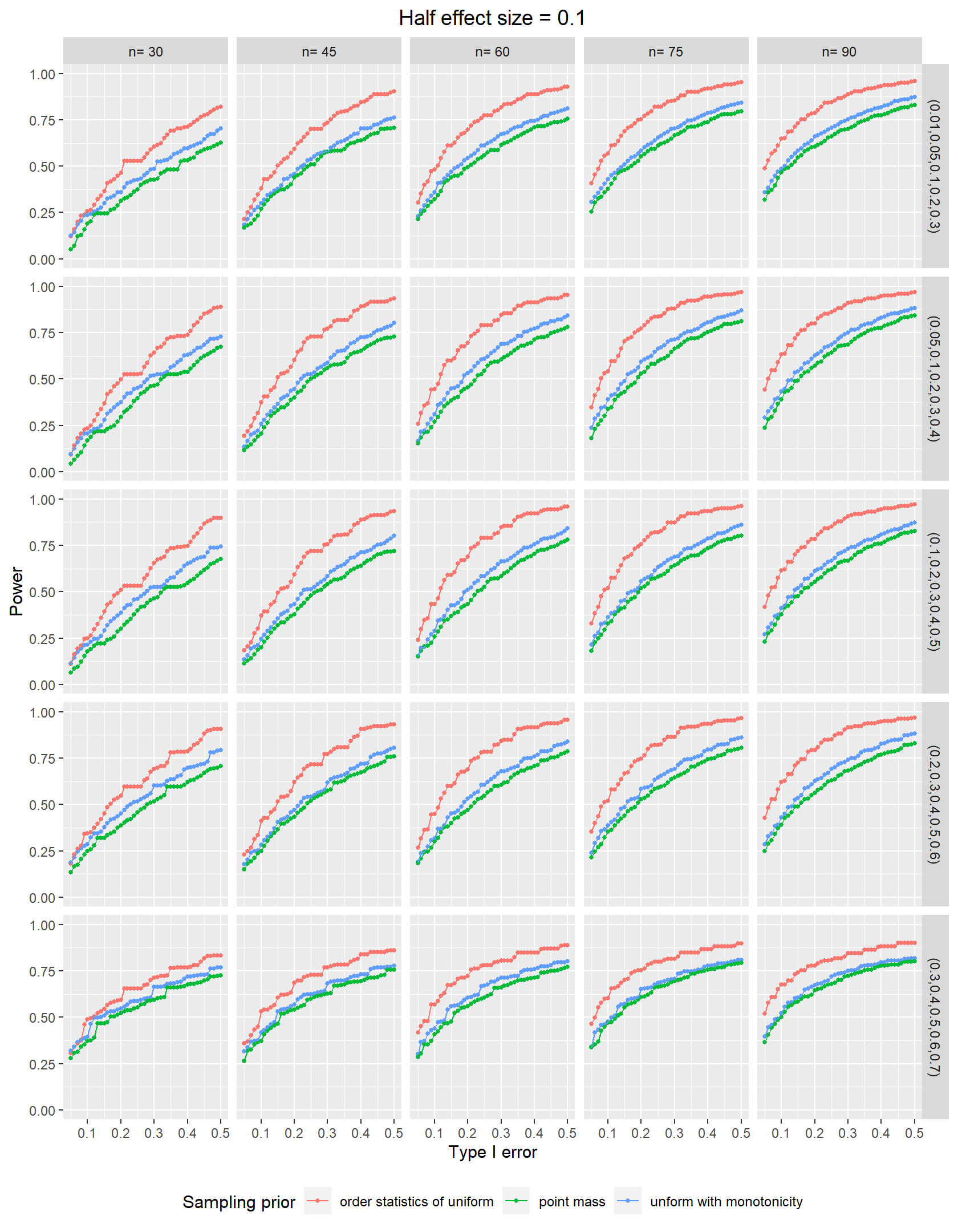}	
	\caption{Power calculated for each combination of $\bp_1^*$ and candidate sample size when the Type I error rate ranges between $5\%$ to $50\%$ and half effect size $\epsilon_1=\epsilon_2=0.1$. Each row indexes an alternative scenario for $\bp_1^*$ and each column indexes a candidate sample size. In each sub-figure, each color indexes a different configuration for $\pi_0^{(s)}(\bp \mid H_0)$.}		
	\label{fig:sampling-1}
\end{figure}

\begin{figure}
	\centering \includegraphics[width=0.9\textwidth]{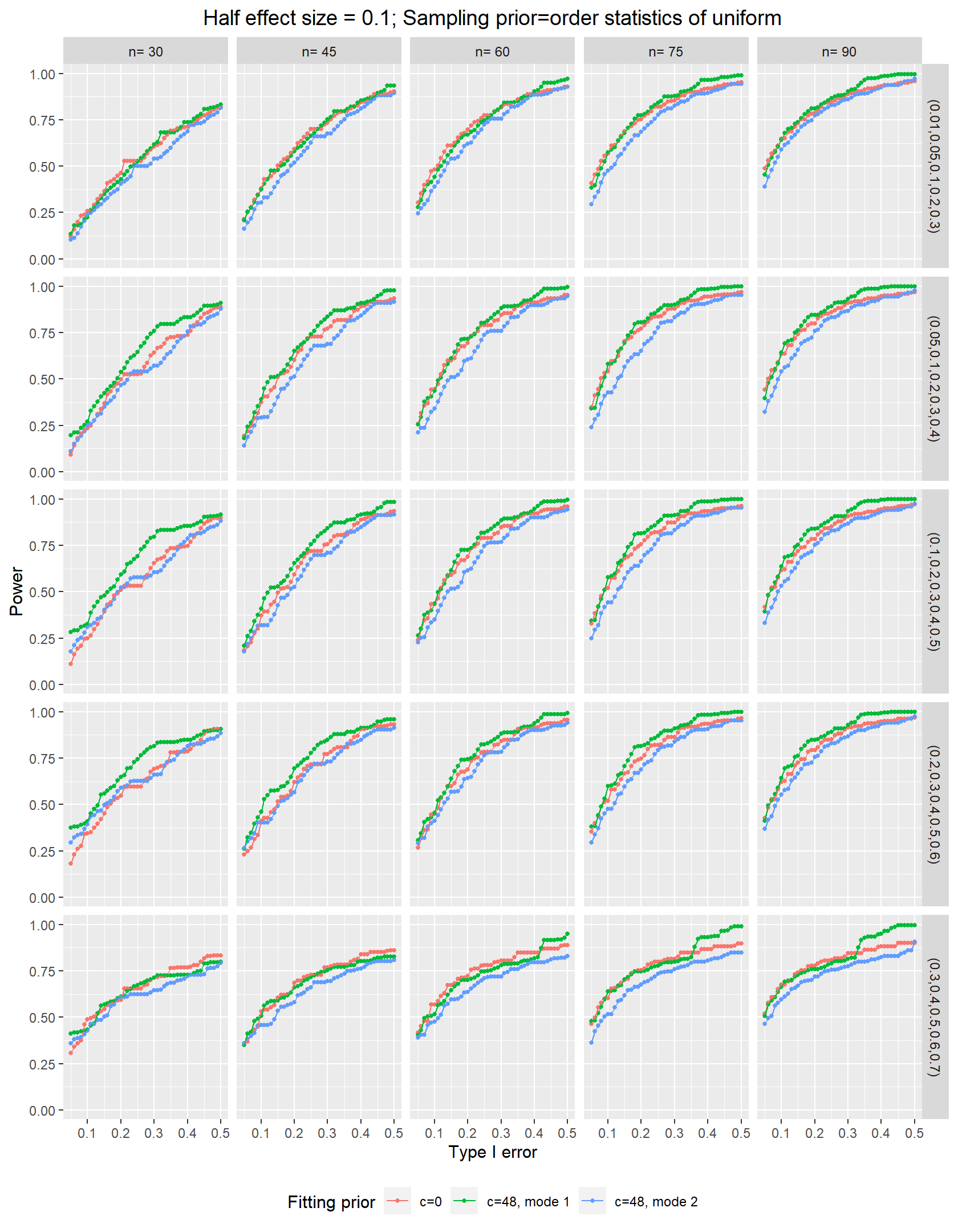}	
		\caption{Power calculated for each combination of $\bp_1^*$ and candidate sample size when the Type I error rate ranges between $5\%$ to $50\%$, half effect size $\epsilon_1=\epsilon_2=0.1$ and $\pi_0^{(s)}(\bp \mid H_0)$ fixed. Each row indexes an alternative scenario for $\bp_1^*$ and each column indexes a candidate sample size. In each sub-figure, each color indexes a different configuration for $c$, $\bm{q^{1j}}$ and $\bm{q^{0j}}$.}			
	\label{fig:cmode-1}
\end{figure}

\begin{figure}
	\centering \includegraphics[width=0.9\textwidth]{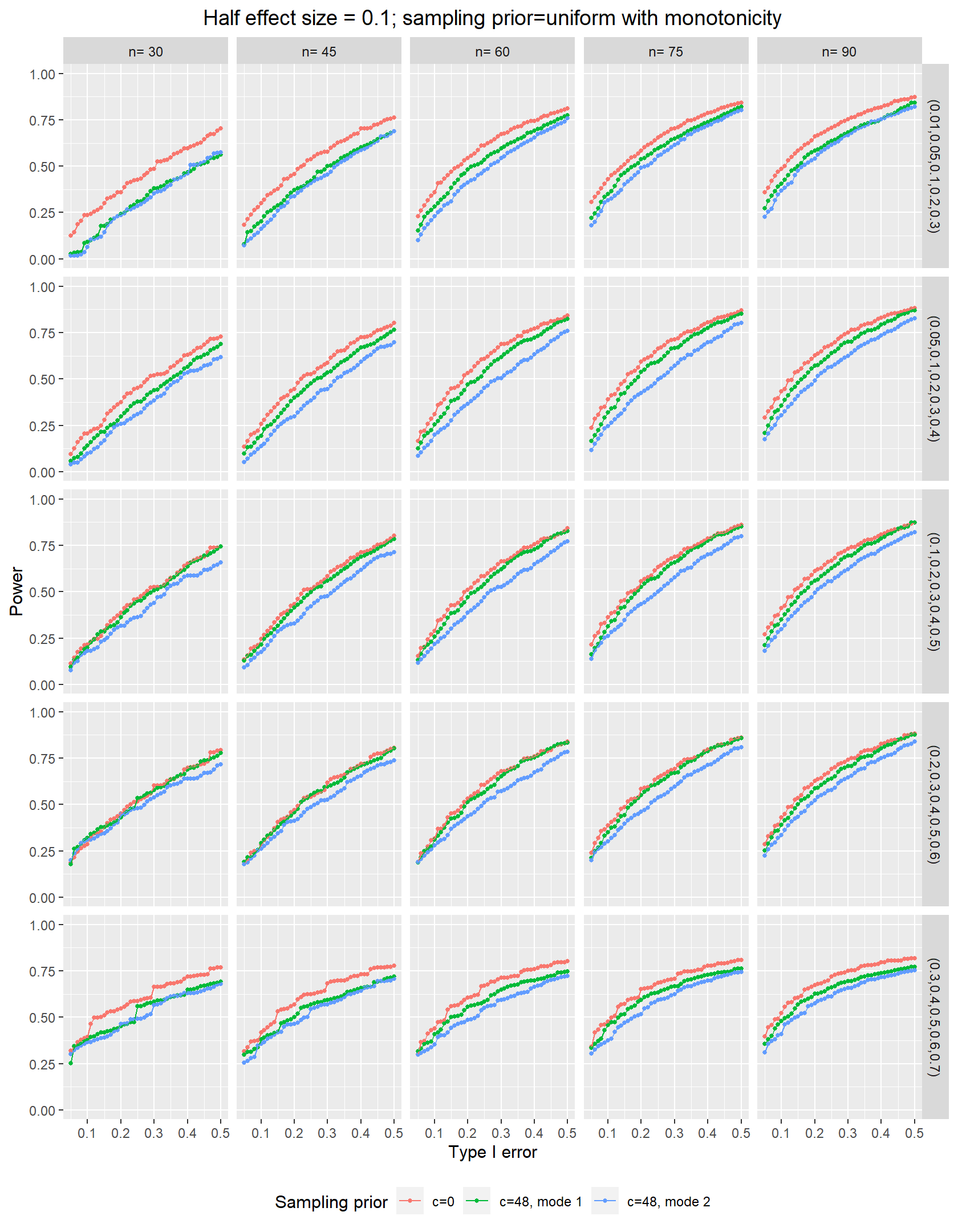}
			\caption{Power calculated for each combination of $\bp_1^*$ and candidate sample size when the Type I error rate ranges between $5\%$ to $50\%$, half effect size $\epsilon_1=\epsilon_2=0.1$ and $\pi_0^{(s)}(\bp \mid H_0)$ fixed. Each row indexes an alternative scenario for $\bp_1^*$ and each column indexes a candidate sample size. In each sub-figure, each color indexes a different configuration for $c$, $\bm{q^{1j}}$ and $\bm{q^{0j}}$.}	
	\label{fig:cmode-2}
\end{figure}

For sensitivity analysis, three types of the sampling priors under $H_0$ (described in Section \ref{Sec:sp}) are used, with $c=0$ (vague prior) and $c=48$ (informative prior). Here, the half effect size is fixed at 0.1. Results are illustrated in Figures \ref{fig:sampling-1}, \ref{fig:cmode-1} and \ref{fig:cmode-2}. \\

\vskip 0.1in
\noindent
Figure \ref{fig:sampling-1} indicates that the specification of the sampling prior under $H_0$ has large impact on the sample size determination. The order statistics of the uniform distribution has the largest power among all situations, uniform with monotonicity comes second, and the point mass has the smallest power given a candidate sample size.  The order statistics of the uniform distribution is the most dispersed prior across the dose range from Figure \ref{fig:sp0} and as a result, more doses are further away from $p_T$. Therefore, it is easier to accept $H_0$ when $H_0$ is true, and the calibrated $BF_0$ from step 1 is larger. \\ 

\vskip 0.1in
\noindent
Figures \ref{fig:cmode-1} and \ref{fig:cmode-2} indicate that, the effects of dispersion parameter $c$ and mode vector $\bm{q^{1j}}$ and $\bm{q^{0j}}$ are generally small and dependent on the sampling prior $\pi_0^{(s)}(\bp \mid H_0)$ and $\bp_1^*$. Here, we investigated two configurations of the mode vector: in Mode 1, we have  $(a_1,a_2,a_3,a_4)=(0.6,0.9,1.05,1.2)$, and in Mode 2,  $(a_1,a_2,a_3,a_4)=(0.3,0.5,1.2,1.5)$. When the order statistics of the uniform is used as the sampling prior under $H_0$, different configurations of $c$, $\bm{q^{1j}}$ and $\bm{q^{0j}}$ give similar results. However, when the uniform with monotonicity is used as the sampling prior, the vague fitting prior when $c=0$ provides larger power than when $c=48$. The fitting prior with mode 1 has larger power than mode 2, which is expected due to its similarity with the $\bp_1^*$ scenario. 

\section{Practical Application for $p_T=0.2$} \label{Sec:otherapp} 
\begin{table}[H]
	\centering
	\caption{For each combination of effect size, Type I error rate and candidate sample size, the range of the power given $p_T=0.2$ is calculated using five different scenarios for $\bp_1^*$ where in each scenario, the true MTD locates at each different dose level.} \label{tab:effectsize0.2}
	\begin{tabular}{c c c c c c c}
		\hline \hline
		\multirow{2}{*}{Half effect size}&\multirow{2}{*}{Type I error}&\multicolumn{5}{c}{Power range}\\
		&&n=30&n=45&n=60&n=75&n=90\\
		\hline
		0.025&0.05&0.09$\sim$0.20&0.11$\sim$0.19&0.10$\sim$0.18&0.11$\sim$0.19&0.11$\sim$0.20\\
		&0.1&0.17$\sim$0.31&0.19$\sim$0.29&0.20$\sim$0.29&0.22$\sim$0.28&0.23$\sim$0.30\\
		&0.2&0.34$\sim$0.45&0.35$\sim$0.41&0.37$\sim$0.45&0.40$\sim$0.45&0.44$\sim$0.49\\
		&0.3&0.43$\sim$0.54&0.48$\sim$0.56&0.50$\sim$0.57&0.55$\sim$0.64&0.58$\sim$0.64\\
		&0.4&0.55$\sim$0.65&0.58$\sim$0.67&0.64$\sim$0.73&0.65$\sim$0.74&0.68$\sim$0.76\\
		&0.5&0.65$\sim$0.76&0.68$\sim$0.77&0.71$\sim$0.80&0.74$\sim$0.82&0.76$\sim$0.82\\
		\hline0.05&0.05&0.12$\sim$0.25&0.11$\sim$0.22&0.11$\sim$0.21&0.15$\sim$0.27&0.18$\sim$0.30\\
		&0.1&0.19$\sim$0.39&0.21$\sim$0.34&0.25$\sim$0.37&0.29$\sim$0.40&0.35$\sim$0.44\\
		&0.2&0.32$\sim$0.47&0.39$\sim$0.49&0.47$\sim$0.56&0.52$\sim$0.55&0.57$\sim$0.62\\
		&0.3&0.46$\sim$0.60&0.52$\sim$0.59&0.62$\sim$0.66&0.66$\sim$0.72&0.69$\sim$0.75\\
		&0.4&0.59$\sim$0.70&0.66$\sim$0.75&0.72$\sim$0.79&0.75$\sim$0.81&0.77$\sim$0.85\\
		&0.5&0.68$\sim$0.79&0.76$\sim$0.84&0.77$\sim$0.84&0.78$\sim$0.87&0.84$\sim$0.90\\
		\hline0.1&0.05&0.20$\sim$0.43&0.23$\sim$0.42&0.36$\sim$0.52&0.46$\sim$0.59&0.53$\sim$0.62\\
		&0.1&0.27$\sim$0.49&0.39$\sim$0.58&0.53$\sim$0.64&0.61$\sim$0.69&0.68$\sim$0.74\\
		&0.2&0.49$\sim$0.66&0.61$\sim$0.74&0.71$\sim$0.77&0.78$\sim$0.83&0.78$\sim$0.85\\
		&0.3&0.64$\sim$0.72&0.72$\sim$0.80&0.78$\sim$0.88&0.80$\sim$0.91&0.81$\sim$0.91\\
		&0.4&0.72$\sim$0.79&0.79$\sim$0.90&0.84$\sim$0.93&0.85$\sim$0.94&0.89$\sim$0.96\\
		&0.5&0.77$\sim$0.86&0.84$\sim$0.93&0.89$\sim$0.97&0.98$\sim$1.00&0.97$\sim$1.00\\
		\hline\hline
	\end{tabular}
\end{table}

\end{appendices}

\end{document}